\documentclass[preprint]{aastex}

\shortauthors{Manset and Bastien}
\shorttitle{Periodic polarimetric observations of PMS binaries}

\begin{document}

\title{Polarimetric variations of binary stars. IV. Pre-main-sequence
spectroscopic binaries located in Taurus, Auriga, and Orion} 
\author{N. Manset\altaffilmark{1} and P. Bastien} \affil{D\'epartement
de Physique, Universit\'e de Montr\'eal,  C.P. 6128, Succursale
Centre-Ville, Montr\'eal, QC, H3C 3J7, Canada, and Observatoire du Mont
M\'egantic} \email{manset@cfht.hawaii.edu, bastien@astro.umontreal.ca}

\altaffiltext{1}{Now at: Canada-France-Hawaii Telescope Corporation,
65-1238 Mamalahoa Hwy, Kamuela, HI 96743, USA} 

\begin{abstract}
We present polarimetric observations of 14 pre-main-sequence (PMS)
binaries located in the Taurus, Auriga, and Orion star forming
regions. The majority of the average observed polarizations are below
0.5\%, and none are above 0.9\%. After removal of estimates of the
interstellar polarization, about half the binaries have an {\it
intrinsic} polarization above 0.5\%, even though most of them do not
present other evidences for the presence of circumstellar dust. Various
tests reveal that 77\% of the PMS binaries have or possibly have a
variable polarization. LkCa~3, Par~1540, and Par~2494 present detectable
periodic and phase-locked variations. The periodic polarimetric
variations are noisier and of a lesser amplitude ($\sim$0.1\%) than for
other types of binaries, such as hot stars. This could be due to
stochastic events that produce deviations in the average polarization, a
non-favorable geometry (circumbinary envelope), or the nature of the
scatterers (dust grains are less efficient polarizers than
electrons). Par~1540 is a Weak-line T~Tauri Star, but nonetheless has
enough dust in its environment to produce detectable levels of
polarization and variations. A fourth interesting case is W~134, which
displays rapid changes in polarization that could be due to eclipses. We
compare the observations with some of our numerical simulations, and
also show that an analysis of the periodic polarimetric variations with
the Brown, McLean, \& Emslie (BME) formalism to find the orbital
inclination is for the moment premature: non-periodic events introduce
stochastic noise that partially masks the periodic low-amplitude
variations and prevents the BME formalism from finding a reasonable estimate
of the orbital inclination.

\end{abstract}


\section{Introduction}
Pre-main-sequence (PMS) stars are objects still contracting to the main
sequence (MS) and are divided into 2 classes according to their masses:
T~Tauri stars (TTS) are low-mass PMS stars with $0.5\;$M$_{\sun}
\lesssim\;$M$\lesssim 2.0\;$M$_{\sun}$, while Herbig~Ae/Be (HAeBe) stars
represent their higher-mass counterparts with 2~M$_{\sun} \lesssim$ M
$\lesssim 10$ M$_{\sun}$. These 2 classes of objects exhibit properties
characteristic of their youth: (1) association with dark or bright
nebulosities (remnant of the parent cloud), (2) emission line spectrum
(spectral type F or later for the TTS, A or B for the HAeBe), (3) IR
excesses, (4) position above the MS in the HR diagram, and (5) presence
of the \ion{Li}{1} absorption line. Both types of PMS stars show signs
of circumstellar (CS) material in their environment, in the form of jets
and bipolar flows, emission excesses in the IR, mm and sub-mm domains,
P~Cygni or inverse P~Cygni profiles, linear polarization, and resolved
disks or envelopes around some of the objects. In the case of binary
stars, the disks can be found around each component (circumstellar (CS)
disks) and/or around the binary system (circumbinary (CB)
disks). Variability is observed in photometry, spectroscopy, and
polarimetry.

TTS are further divided into 2 classes according to the width of their
H$\alpha$ emission line. Classical TTS (CTTS) have wide H$\alpha$
emission lines with $W_{\lambda}> 5$ \AA\, whereas the Weak-line TTS
(WTTS) have $W_{\lambda}<5$ \AA\ (Bertout 1989). Naked TTS (NTTS) are a
subclass of the WTTS (Walter 1986; Wolk \& Walter 1996), although some
tend to use both terms as synonyms. Indeed, in addition to narrower
H$\alpha$ lines, the NTTS specifically do not show evidence for
circumstellar (CS) material in their environments in the form of IR
excesses, whereas the WTTS may or may not (Wolk \& Walter
1996). WTTS and NTTS should not be confused with post-T~Tauri stars,
which are still above the MS, but more evolved than TTS. For
a review on T~Tauri stars and Herbig~Ae/Be stars, see for example
Bertout (1989) and Catala (1989).

Dust grains produce polarization by scattering, polarization that has
been known for a number of years (see Bastien 1996 for a review). In
general, the red part of the visible spectrum of TTS exhibits linear
polarizations of 1--2\%, but the polarization distributions of CTTS and
WTTS are markedly different: there are a few CTTS which have a very high
polarization (more than 5\%, sometimes up to 15\%), but WTTS almost all
have polarization levels below 2\% (Bastien 1982, 1985; M\'enard \&
Bastien 1992). Also, most TTS that have an active CS disk have higher
polarization and near IR excess than other TTS (Yudin 2000). On average,
HAeBe stars are more polarized than TTS (3.0\% versus 1.6\%), and there
are clear differences between the polarization distributions for TTS and
HAeBe stars (Yudin 2000). Most PMS stars have statistically higher
polarization levels than more evolved stars which are closer to the MS
and there is clear evidence of changes in polarimetric behavior
of stars during the evolution from PMS to MS star; these changes in
polarimetric behavior are related to the evolution of the CS environment
(Yudin 2000).

The IR color indices and color excesses of TTS are well correlated with
the polarization (Bastien 1985), implying that the same grains are
responsible for both phenomenons. A compilation of polarization data for
almost 500 TTS and HAeBe stars studied by Yudin (2000) reveals that for
85\% of the sample stars, there is a correlation between the degree of
polarization and the IR color index $(V - L)_{\rm obs}$ and the color
excess $E(V-L)$. This polarization is variable in a majority of objects:
more than 85\% of the TTS, and more than 70\% of HAeBe stars present
variability (Bastien 1988; M\'enard \& Bastien 1992). Variations are
sometimes large and fast ($\Delta P > 0.5$\%, $\Delta \theta >
15\arcdeg$, within 5 days, Bastien 1985), sometimes associated with
luminosity and color changes (Grinin et al. 1991; Grinin, Kolotilov, \&
Rostopchina 1995), and in some cases periodic.

We have shown (Manset \& Bastien 2000, 2001a, hereafter Paper~I and II)
how the linear polarization of a binary surrounded by circumstellar
matter varies periodically as a function of the orbital period, and how
the geometry of the disks, the nature and characteristics of the
scatterers, the masses of the stars, and the orbit characteristics
affect the polarimetric curves. Models can be used to find the orbital
inclination from observed polarimetric variations (see for example Rudy
\& Kemp 1978; Brown, McLean, \& Emslie 1978). The work from Brown et
al. (1978) (hereafter BME) uses first- and second-order Fourier analysis
of the Stokes curves to give, in addition to the orbital inclination,
moments related to the distribution of the scatterers in the CS and CB
environments. The BME formalism was developed for Thomson scattering in
optically thin envelopes, and for binaries in circular orbits. Since
polarization in PMS stars is produced by scattering on dust grains, and
most of the known spectroscopic PMS binaries have eccentric orbits, the
BME formalism can not be used a priori. However, we have shown (Paper~I
and II) that the BME analysis can still be applied in those cases, with
a few limitations.

In this context, we have obtained polarimetric observations of 24
spectroscopic PMS binaries. A detailed analysis was presented for one of
these binaries, the HAeBe star MWC~1080 (Manset \& Bastien 2001b,
hereafter Paper~III). Here we report the complete observations and
detailed analysis for the PMS binaries located in the Taurus, Auriga,
and Orion star forming regions (SFRs). Binaries located in the
Scorpius and Ophiucus regions will be analyzed in a future paper.

\section{Observations}
We have observed 24 PMS spectroscopic binaries with $\delta \gtrsim
-25\arcdeg$; the shortest-period and brightest ones ($P \lesssim 35$ d,
$V \lesssim 12.0$) were followed with $\approx 10$ or more
observations. These stars were chosen mainly from the list in Mathieu
(1994), to which we added subsequent discoveries. Tables~\ref{Tab-Coord}
and \ref{Tab-Spectro} present basic information (other names,
coordinates, location), and spectroscopic and orbital data (spectral
type, PMS type, orbital period and eccentricity, orbital inclination
when known, and distance) for the 14 binaries located in the Taurus,
Orion, and Auriga SFRs. 

The binaries were observed at the Observatoire du Mont M\'egantic (OMM),
Qu\'ebec, Canada, between 1994 September and 1999 March, using a
$8\farcs2$ aperture hole and a broad red filter (RG645: 7660 \AA\
central wavelength, 2410 \AA\ full width at half maximum). Polarimetric
data were taken with Beauty and The Beast, a two-channel photo-electric
polarimeter, which uses a Wollaston prism, a Pockels cell, and an
additional quarter-wave plate. The data were calibrated for instrumental
efficiency, instrumental polarization (due to the telescope's mirrors),
and zero point of position angle, using a Glan-Thomson prism,
non-polarized standard stars, and polarized standard stars
respectively. The observational errors were calculated from photon
statistics, and also include uncertainties introduced by the previously
mentioned calibrations. The final uncertainty on individual measurements
of the polarization $P$ is usually in the range 0.03--0.05\%. The
relative errors in position angle $\theta$ can be as low as 0.1\arcdeg,
but due to instrumental effects, systematic errors, and the calibration
procedure itself, the absolute errors on the position angles are of the
order of 1$^\circ$. For more details on the instrument and the
observational method, see Manset \& Bastien (1995, 2001b) and Manset
(2000). Table~\ref{Tab-averpol} summarizes the observations and gives
the number of observations, the average polarization and position angle.

\section{Estimation of the interstellar polarization}
Polarimetric observations are usually a sum of interstellar (and
sometimes also intra-cluster) and intrinsic polarizations. An estimation
of the interstellar (IS) polarization for each object observed can be
used to assess the presence of intrinsic polarization. We have used the
Heiles (2000) catalog of over 9000 polarization measurements to
determine if the observed polarization for the PMS binaries studied here
is of intrinsic or IS origin, or a combination of both. This catalog is
an improvement over the one from Mathewson et al. (1978): it contains
additional observations, all data were verified, and more precise
coordinates are given.

For each observed PMS binary, the catalog was scanned to select at least
20 close stars {\it with a similar distance}. Depending on the stellar
density and number of measurements in the catalog, this led to the
selection of a region between 1 and 15\arcdeg\ in radius around the
target, and within 70 to 350~pc of it. The stars selected from the
catalog are used to give an average of the IS polarization in that
region around the target, and also to find the ratio
$P/E(B-V)$. Finally, based on an extinction value for our target and
assuming this extinction is of IS origin only, an estimate of the IS
polarization for the target is calculated, along with the average IS
polarization angle. This angle is calculated with a simple average and
with a distance-weighted average of the polarization angles of all the
stars selected. In all the cases here, the two values are similar to
within their uncertainty, which indicates that the alignment is good
over all the regions studied, and the IS polarization value estimated is
reliable. If the position angles for the IS polarization and for the
target are different, it points to an intrinsic origin for at least part
of the polarization measured.  Intrinsic polarization is also deduced
from polarimetric variability.

Results are presented in Table~\ref{Tab-averpol}, where the weighted
averages of the observed $P$ and $\theta$ are given along with the
possible origin of the observed polarization: a $\star$ symbol indicates
intrinsic polarization while IS stands for interstellar
polarization. When IS comes before a $\star$ symbol, the IS component of
the polarization is probably stronger than the intrinsic one, and vice
versa. The following columns present our calculation of the IS
polarization (polarization and position angle, along with their
uncertainties), based on a weighted average of the polarization of
neighboring stars located at similar distances, where more weight is
given to stars closer to the PMS binary. Since non-weighted and weighted
averages give similar results, only the latter average is presented. We
also give $N_{IS}$, the number of measurements used to estimate this IS
polarization, and the radius and the interval of distance of the region
considered. Subtracting the IS polarization from the observed one gives
the intrinsic polarization, shown in the last columns.

Note that the IS polarization given in columns 6--9 of
Table~\ref{Tab-averpol} is only an {\it estimate} for the {\it whole
region} around a binary, and in some cases might not apply to a given
binary. In particular, it may not include very localized intra-cluster
polarization. Consequently, the intrinsic polarizations given in the
last columns should be considered crude estimations only, intended to
give an idea of the polarimetric characteristics of the observed
binaries as a whole, and not a definitive value of the intrinsic
polarization for each binary.

To determine if an observed polarization has an IS component, we give
more weight to the value of $\theta_{\rm IS}$ deduced from neighboring
stars than to $P_{\rm IS}$. We do not have the wavelength dependence of
the observed polarization, which can usually be used to extract the IS
component. To help determine if there is an intrinsic component of
polarization, we also use the level of polarimetric variability since IS
polarization is very stable.

\section{Polarimetric variability \label{p3-sec-var}}
Since a majority of single PMS stars are variable polarimetrically
(Bastien 1982; Drissen, Bastien, \& St-Louis 1989; M\'enard \& Bastien
1992), we expected PMS binaries to also be polarimetrically 
variable, either periodically or not.

\subsection{Variability tests}
We applied various tests to check the polarimetric variability or
stability of PMS binaries: minimum and maximum values, variance test,
$Z$ test and a similar one we call $\sigma_1$ and $\sigma_2$ test, and
finally, a $\chi^2$ test. We have not retained skewness and kurtosis
tests (which measure departure from a Normal distribution) because the
number of data points is for some stars very limited, and the results
sometimes difficult to interpret. It should be noted that these tests
usually assume that the parent distribution of the quantity measured
(here, $P$, $\theta$, or the Stokes parameters $Q$ and $U$) is
Normal. Since $P$ and $\theta$ are not distributed Normally (Serkowski
1958), these tests should in general be applied only to the Stokes
parameters $Q$ and $U$.

Many PMS binaries show observations with polarization levels and/or
position angle well below or above the bulk of the data. Assuming these
observations are atypical but nonetheless real, we have removed them
before using the tests. This allows us to study the more typical
variations. We believe these atypical observations are real: close
examination of polarization observations taken over 5 years of
non-polarized standard stars (84 observations), polarized standard stars
(53 observations), 3 stars that were followed for many consecutive hours
(121 observations) did not show odd observations like the ones we
repeatedly saw for PMS binaries (Manset 2000). Therefore, we believe
these observations were due to some eruption-like events or significant
modifications in the CS environment (e.g., formation/destruction of
condensations, accretion events), and not because of an instrumental
problem.

\subsection{Maximum and minimum values}
One crude but easy way to check for variability in a set of observations
is to compare the difference between the maximum and minimum values of a
quantity with its average or typical observational uncertainty; variable
observations will have maximum and minimum values well outside the range
expected from the observational uncertainty. This test was applied to
all the sets of observations, after removing the observations that
showed the most deviation from the majority of the observations. This
test should not be used alone since it does not take into account the
entirety of the data.

\subsection{Variance test}
Given a set of $N$ observations $D_i$ where $i=1, ..., N$, we can 
calculate the variance of the sample (which is a measure of the 
``width'' of the observations, or of the scatter from the mean, or of
the ``variability'' around a central value, usually measured with the
mean $\overline{D}$) with the formula: 
\begin{equation}
\sigma^2_{\rm sample} = \frac{1}{N-1} \sum_{i} (D_i - \overline{D})^2,
\end{equation}
which, in computer programs, can also be calculated more rapidly (without
having first to determine the mean $\overline{D}$) with:
\begin{equation}
\sigma^2_{\rm sample} = \frac{\sum_{i} D_i^2}{N-1} - 
\frac{(\sum_{i} D_i)^2}{N(N-1)} .
\end{equation}
This sample variance can then be compared to the standard deviation of 
the mean, which gives the error from photon statistics as if all the
observations had been added together:
\begin{equation}
\sigma^2_{\rm mean} = \frac{1}{\sum_{i} 1/\sigma_i^2}.
\end{equation}

For a set of observations of a non-variable quantity, the sample variance
will be low (the observations are all clustered closely to the mean) and
similar to the standard deviation of the mean. But if there is
variability, the ``width'' of the observations will be greater than the
standard deviation of the mean: $\sigma_{\rm sample} \ge \sigma_{\rm mean}$.
It should be noted that it is assumed the parent population is distributed
Normally, so this test will be meaningful only for the Stokes parameters.

\subsection{$Z$ test}
Given a quantity $D$ for which we have $N$ measurements $D_i \pm
\sigma_i$, we can calculate the weighted mean:
\begin{equation}
\overline{D}_w = \frac{\sum_{i} (D_i / \sigma^2_i)}{\sum_{i} (1/\sigma^2_i)},
\end{equation}
with its associated variance, which we will call the ``external'' variance:
\begin{equation}
\sigma^2_{\rm w1} = \frac{1}{\sum_{i} 1/\sigma^2_i}.
\end{equation}
Alternatively, the variance may be computed according to the weighted 
residuals, giving an ``internal'' assessment of the distribution:
\begin{equation}
\sigma^2_{\rm w2} = \frac{\sum_{i} \frac{(D_i - \overline{D}_w)^2}{\sigma^2_i}}
{(N-1) \sum_{i} 1/\sigma^2_i}.
\end{equation}

We then build the quantity $Z$:
\begin{eqnarray}
Z &=& \frac{\sigma_{\rm w2}}{\sigma_{\rm w1}}\\
  &=& \sqrt{\frac{\sum_{i} \frac{(D_i - \overline{D}_w)^2}{\sigma^2_i}}{(N-1)}}.
\
\
\end{eqnarray}

If the data are ``well behaved'', $Z$ should equal unity (Brooks et al. 1994). 

The standard error of $Z$ is (Topping 1972):
\begin{equation}
\sigma_Z = \frac{1}{2(N-1)}.
\end{equation}

This test should be applied to $Q$ and $U$, and not to $P$ and $\theta$.
If $Z\approx1$ within its standard error, then the measurements are
``consistent'', and there is no variability. If $Z$ differs
significantly from 1, then there may be variability. For details, see
Brooks et al. (1994).

\subsection{$\sigma_1$ and $\sigma_2$ test}
The $\sigma_1$ and $\sigma_2$ test is similar to the $Z$ test, except
that it is applied to the polarization $P$ and its position angle
$\theta$. 
Given a set of $N$ polarimetric observations of the Stokes parameters 
$Q$ and $U$,
we can compare the variance of the polarization $\sigma^2_1(P)$ 
with the scatter from the mean of the polarization $\sigma^2_2(P)$:
\begin{equation}
\sigma^2_1(P) = N \left[ \sum_{i} \frac{1}{\sigma^2_i} \right]^{-1},
\end{equation}
\begin{equation}
\sigma^2_2(P) = \frac{\sum_{i} (Q_i - {\overline{Q}})^2 + \sum_{i} (U_i
- {\overline{U}})^2} {2(N-1)}.
\end{equation}

For the position angle, we can calculate $\sigma_1(\theta)$ and 
$\sigma_2(\theta)$ with the following formula:
\begin{equation}
\sigma(\theta) = 28\fdg65\,\frac{\sigma(P)}{P}. 
\end{equation}

If $\sigma_2 > \sigma_1$ then there may be variability. Bastien et al.
(1988) used this method to assess the polarimetric variability of
polarimetric standard stars. Clarke \& Naghizadeh-Khouei (1994) make the
remark that $\sigma_2$ should be the weighted mean.

\subsection{$\chi^2$ test}
In Bastien et al. (1988) and Bastien (1982), a $\chi^2$-based
method is presented: $\chi^2$ values are calculated for $Q$ and $U$
separately, using 1$\sigma_i$ and $1.5\sigma_i$. Then, the
probability to obtain a given value of $\chi^2$ in a Gaussian
distribution is found for each of the four $\chi^2$ values. The star is
variable if at least 2 of the four $\chi^2$ values are over 0.95; the
star is suspected to be variable if one out of four $\chi^2$ values is
over 0.95. 

\subsection{How to use the variability tests}
The minimum and maximum values test is not a robust test; stars that do
not show minimum and maximum values out of the range expected from the
observational uncertainty ($\pm 1 \sigma$ intervals) are found to have a
non-variable behavior by other tests, but the converse is not true.

The variance and $Z$ tests were used only for the Stokes parameters (and
not for the polarization and its position angle). The $Z$ test was
considered positive if $Z > 1.0 + \sigma_Z$; values of $Z$ below
($1.0\;-\;\sigma_Z$) were considered dubious, since $Z < 1.0$ means the
data are too well behaved with respect to the statistical
(observational) uncertainty. The variance test was considered positive
if $\sigma_{\rm sample} \geq 3.0\, \sigma_{\rm mean}$; that way, the
variance and $Z$ tests would give consistent results. With $\sigma_{\rm
sample} \geq \sigma_{\rm mean}$, the variance test would sometimes see
variability where the $Z$ test would not.

The $\sigma_1$ and $\sigma_2$ tests were not considered as primary
tests, since they apply to $P$ and $\theta$, which are not Normally
distributed, and because $\sigma_2$ is not weighted. But the results of
this test agree with the conclusions reached with other tests.

For the $\chi^2$ test, as in Bastien et al. (1988) and Bastien (1982),
the star is considered variable if at least 2 of the four $\chi^2$
values are over 0.95; the star is suspected to be variable if one out of
four $\chi^2$ values is over 0.95.

Using, for $Q$ and $U$ separately, the $\sigma_{\rm sample} \geq 3.0\,
\sigma_{\rm mean}$ test, the $Z > 1.0 + \sigma_Z$ test, and the $\chi^2$
test, we established the variability, suspected variability, suspected
stability, and stability with the following criteria (see also
Table~\ref{Tab-UseVarTests}): 
\begin{itemize}
\item To be considered variable, a star must have at least 2 positive
results from the $Z$ and variance tests and at least 2 positive
results from the $\chi^2$ test; 
        \begin{itemize}
        \item if there are only 2 positive results from the $Z$ and 
variance tests, they must be 2 positive $Z$ tests, or a positive $Z$ and
positive variance tests for the same Stokes parameter; 
        \item if there are only
2 positive results from the $\chi^2$ test, they must be for the same
Stokes parameters as the positive $Z$ and variance tests. 
        \end{itemize}
\item To be considered as suspected variable, the results of the $Z$ and
variance tests must be the same as for the variable conclusion, but only
one $\chi^2$ value must be positive. 
\item A star is considered as possibly constant if the $\chi^2$ test is
negative (for the 4 $\chi^2$ values), but at least 3 values from the $Z$
and variance tests indicate possible variability. 
\item If the $Z$ and $\chi^2$ tests are negative and only 1 or 2 values
of the variance test indicate variability, the star is considered to be
constant. 
\end{itemize}

\subsection{Results of variability tests for the PMS binaries}
Table~\ref{Tab-AmpVar} presents the amplitude of the polarimetric
variations, calculated by simply taking the difference between the
minimum and maximum values of $P$, $\theta$, $Q$ and $U$; note that for
some cases, very atypical observations were not considered. The details
of the variability tests presented in this sections are shown in
Table~\ref{Tab-VarDetails}, where we give for each star the number of
observations used for the variability tests, $\sigma_{\rm sample}$ and
$\sigma_{\rm mean}$, $Z$ and its standard error, and $P{\chi^2}$,
calculated with $1\sigma$ and $1.5\sigma$. The conclusions of the
variability tests are shown in Table~\ref{Tab-Var}, where we have
classified the stars as ``variable'', ``suspected variable'', and
``constant''. Of the 12 binaries which were tested for variability, 42\%
(5/12) are statistically variable (V773~Tau, V826~Tau, GW~Ori, Par~1540,
W~134), and 58\% (7/12) are variable or possibly variable (Par~2486,
Ori~569). DQ~Tau and UZ~Tau E/W could not be tested because of the low
number of observations ($N=1$ and $N=2$ respectively). Five binaries
have statistically constant polarization: LkCa~3, NTTS~045251+3016,
Ori~429, Par~2494, and VSB~126. We confirm the variability already
reported for V773~Tau (M\'enard \& Bastien 1992). Our additional data
establish the variability for the previously suspected variables V826~Tau
(M\'enard \& Bastien 1992) and GW~Ori (Bastien 1985).

\section{Periodic polarimetric variations \label{p3-sec-pervar}}
In addition to the general variability, which is a known property of
single PMS stars, PMS binaries will also present periodic polarimetric
variations caused by the orbital motion, even if in some cases the
amplitude may be too small to be detected with the currently available
instruments or masked by non-periodic or pseudo-periodic variations. In
the case of Mie scattering, we have also shown in Paper~II that dust
grains, which are mostly responsible for the polarization here, are less
efficient polarizers and produce smaller amplitude variations than
electrons. This is an indication that periodic polarimetric variations
could be more difficult to observe in PMS binaries than in, for example,
hot binaries surrounded by electrons, which can easily exhibit
variations of a few tenths of a percent (see for example Robert et
al. 1990; Robert et al.  1992). The size of the grains also determines
the amplitude of the polarimetric variations: dust grains with radii
$\sim 0.1 \micron$ produce the largest polarimetric variations
(Paper~II).

To look for periodic polarimetric variations, the known orbital periods
are used to calculate the orbital phase for each observation of each
star. The polarization $P$, its position angle $\theta$, and the Stokes
parameters $Q$ and $U$ are plotted as a function of the orbital phase
(see Figures~\ref{Fig-v773tau} to \ref{Fig-vsb126}). When enough data
are available, observations are represented as first and second
harmonics of $\lambda=2\pi\phi$, where $\phi$ is the orbital phase:\\
\begin{eqnarray}
Q &=& q_0 + q_1 \cos \lambda + q_2 \sin \lambda + q_3 \cos 2\lambda +
q_4 \sin 2\lambda, \label{p3-eq-qfit}\\
U &=& u_0 + u_1 \cos \lambda + u_2 \sin \lambda + u_3 \cos 2\lambda +
u_4 \sin 2\lambda. \label{p3-eq-ufit}
\end{eqnarray}
The coefficients of this fit are then used to find the orbital
inclination, following the BME formalism and using the first or second
order Fourier coefficients, although it is usually expected that second
order variations will dominate (BME):\\
\begin{eqnarray}
\left[ \frac{1-\cos i}{1+\cos i} \right]^2 &=& \frac{(u_1+q_2)^2 +
(u_2-q_1)^2}{(u_2+q_1)^2 + (u_1-q_2)^2} \label{EQ-iO1-p3} , \\
\left[ \frac{1-\cos i}{1+\cos i} \right]^4 &=& \frac{(u_3+q_4)^2 +
(u_4-q_3)^2}{(u_4+q_3)^2 + (u_3-q_4)^2} \label{EQ-iO2-p3}.
\end{eqnarray}

As can be seen in Figures~\ref{Fig-v773tau} to \ref{Fig-vsb126}, not all
PMS binaries show periodic polarimetric variations, and even then, those
are not always clearly seen, which was expected to some extent.
Periodic polarimetric variations can be caused by the binarity (orbital
motion) or the presence of hot or cool stellar spots, among a few
reasons. Classical T~Tauri Stars (CTTS) are known to have both cool and
hot spots, and WTTS generally have only cool spots (Bouvier et
al. 1993), some of which can be stable over periods of several months
(on V410 Tau for example, Herbst 1989). Since most of the binaries
observed here are WTTS and in general only display small photometric
variations, we believe that the spots causing the photometric
variations, if present, are small, and then have a very small effect on
the polarization. Nonetheless, as can be seen in the figures, the
appearance and disappearance of transient spots could be the cause of
the non-periodic variations that introduce some scatter about the
periodic variations. In addition to stellar spots, non-periodic
phenomenon such as eruptive events, variable accretion and
rearrangements of the CS or CB material can cause pseudo-periodic
polarimetric variations that may mask the strictly periodic ones,
especially if the observations are taken over many orbital periods, as
is the case here.

Despite these difficulties, some binaries present periodic
variations. To investigate the significance of this periodicity, a Phase
Dispersion Method (Stellingwerf 1978) and a Lomb normalized periodogram
algorithm (Press et al. 1997) were used. The Phase Dispersion Method
(PDM) is a least-squared fitting technique suited for non-sinusoidal
time variations covered by irregularly spaced observations, and finds
the period that produces the least scatter about the mean curve. The
Lomb normalized periodogram (LNP) method is more powerful than Fast
Fourier Transform methods for uneven sampling, but still assumes the
curve is sinusoidal, which may not be always appropriate for the
polarimetric observations presented here. The periods found by using
both methods are very similar to one another for a given star, but the
significance is usually marginal.

\section{Comments on individual stars \label{p3-sec-stars}}
The detailed observations are presented in Tables~\ref{Tab-v773tau} to
\ref{Tab-autres} and in Figures~\ref{Fig-v773tau} to \ref{Fig-vsb126}.


\subsection{V773~Tau = HBC~367 = HD~283446}
V773~Tau is a triple system in which the 51.075-day spectroscopic binary
(Welty 1995) has a projected separation of 0.34~AU (Jensen \& Mathieu
1997). The third star is reported to be at $0\farcs17$ from the
spectroscopic binary at a position angle of 295\arcdeg\ (Leinert et
al. 1993), or $0\farcs112$ at position angle 295\arcdeg\ (Ghez,
Neugebauer, \& Matthews 1993). Given the size of our aperture hole, this
third star is also included in our measurements. Jensen, Mathieu, \& Fuller
(1994) report for the tertiary a projected separation of 16~AU. The IR
excess of $K-N=3.4$~mag indicates the presence of an optically thick
inner disk at 10~\micron\ (Simon \& Prato 1995). No CB disk was detected
at 2.7~mm by Dutrey et al. (1996), but Jensen \& Mathieu (1997) argue
that the submm continuum emission must arise in a CB disk, although the
presence of a third star could mean the submm and IR excesses are not
coming from a CB disk, but from a CS disk around the tertiary. The
$3\sigma$ upper limit to the disk's mass, based on 800~\micron\
observations, is 0.001~M$_{\sun}$ (Jensen et al. 1994). V773~Tau is
highly variable in the mm, submm, and radio domains (Jensen et al. 1994;
Dutrey et al. 1996), but simultaneous observations in the radio,
optical, and X-ray regions by Feigelson et al. (1994) showed variations
only in the radio.

From the polarization catalog of Heiles (2000), we find a very low
IS polarization in the vicinity of V773~Tau, 0.07\% $\pm$
0.08\% at a position angle 72\arcdeg $\pm$ 36\arcdeg\ (see
Table~\ref{Tab-averpol}). This value was found by averaging the
polarization of 24 stars located within 15\arcdeg\ and 85 pc of
V773~Tau. A map of the IS polarization in the vicinity of V773~Tau is
presented in Figure~\ref{Fig-Polis_V773_Lk_UZ}, where it is seen that
the polarization of neighboring stars is low. V773~Tau's polarization
is significantly higher than the IS estimate, and is variable (see
below), pointing to intrinsic polarization. We conclude that V773~Tau's
polarization is intrinsic with maybe a small IS component.

Data are presented in Table~\ref{Tab-v773tau} and
Figure~\ref{Fig-v773tau}. V773~Tau's mean polarization at 7660 \AA\ is
0.35\% at 88\arcdeg. This position angle is not correlated with the
position angle of the third star\footnote{Since position angles on
the sky are measured from 0\arcdeg\ to 360\arcdeg\ but polarization
position angles only from 0\arcdeg\ to 180\arcdeg\, the 295\arcdeg\
position angle for the third star would give a polarization position
angle of 65\arcdeg.}. V773~Tau was observed in polarimetry by Bastien
(1982, 1985) and M\'enard \& Bastien (1992). Combined with the new data,
the earlier observations (taken in red filters, centered on $\approx
7600$ \AA, and having widths of $\approx 800$ \AA\ for the data of
Bastien (1982) and M\'enard \& Bastien (1992), and 2410 \AA\ for our
data) reveal that the polarization and especially its position angle are
variable on a time scale of a decade: the polarization was 0.10\% at
72\arcdeg\ in fall 1978, 0.33\% at 108\arcdeg\ in winter 1985, and
0.35\% at 88\arcdeg\ in 1999. There are also large variations as a
function of wavelength.

Using all data available at the time, M\'enard \& Bastien (1992) found
V773~Tau to be variable at least in position angle, in the blue and
green parts of the optical spectrum. We confirm that $P$ and $\theta$
are variable, in the red spectral domain. This variability is a strong
indication that part of the polarization is intrinsic, in agreement with
the presence of some material as indicated by the IR and mm
observations and with our analysis of the IS polarization in this
region. Our variability analysis is based on 6 observations, one of
which presents a polarization ($P\approx0.2$\%) well below the average
of the 5 other observations ($P\approx0.4$\%), although its position
angle agrees well with the other observations. This low polarization
observation was the first observation taken, and it was taken a year
before the next one. This might indicate a change in the environment of
this triple system over the course of one year.  All the polarimetric
data available present evidence for long-term (decades, years) and
short-term (months) variability, as well as variability as a function of
wavelength. The limited number of observations (6) does not permit us to
find any periodic polarimetric variations or to do a fit according to
equations~\ref{p3-eq-qfit} and \ref{p3-eq-ufit}.

Since the third star in this system (included in our aperture hole) is
faint (brightness ratio at $K$ of 0.13; Leinert et al. 1993), most of
the polarization comes from the spectroscopic binary. As mentioned
before, the submm emission could come from a CS disk around the tertiary
instead of a CB disk around the spectroscopic binary, but such a
geometry would probably not produce the polarimetric variations
observed over a few months.

\subsection{LkCa~3 = HBC~368}
LkCa~3 is a triple system (Simon \& Prato 1995) in which the 12.941-day
period secondary (Mathieu 1994) is $0\farcs47$ from the primary (Dutrey
et al. 1996), at a position angle of 78\arcdeg\ (Leinert et al. 1993;
Ghez et al. 1993). In a series of 26 $BVR$ photometric observations,
Grankin (1993) found photometric variations of the order of 0.7~mag in
$V$, but no period could be found. The low excess $K-N=0.1$ mag
indicates that there might be an optically thin inner disk at
10~\micron\ or no disk at all (Simon \& Prato 1995). Wolk \& Walter
(1996) attribute the IR flux to the photosphere, and not to optically
thin material. No CB disk was detected at 2.7~mm (Dutrey et al. 1996).

LkCa~3 is within half a degree of V773~Tau, in a similar region of low
polarization (see Table~\ref{Tab-averpol} and
Figure~\ref{Fig-Polis_V773_Lk_UZ}). Most if not all of the low
polarization observed for LkCa~3 could therefore be of IS origin, which
agrees with the fact that there is little or no indication of CS or CB
material. Data are presented in Table~\ref{Tab-lkca3} and in
Figure~\ref{Fig-lkca3}. LkCa~3's average polarization, 0.05\% at
76\arcdeg, has a position angle close to the position angle of the
secondary (78\arcdeg). Statistically, the polarization is constant,
which agrees with its predominant IS origin. However, there are some
periodic behavior in the position angle: between phases 0.2 and 0.65,
the variations outline a sinusoidal wave (see
Figure~\ref{Fig-lkca3}). Statistical tests that check for variability do
not taken into account such systematic, although small, variations, and
low-level variability may not be completely ruled out.

\subsection{V826~Tau = HBC~400 = TAP~43}
The projected separation between the two components of this 3.88776-day
binary (Mathieu 1994) is 0.06~AU (Jensen et al. 1994). Although this
WTTS is a PMS star with an estimated age of $10^6$~yr (Mathieu, Walter,
\& Myers 1989), it is a mature system, with a circular orbit and no
evidence for circumstellar material. It shows weak H$\alpha$ emission
lines superimposed on a normal continuum, UV excess, and strong X-ray
emission, but lies above the MS (see Mundt et al. (1983), who review
the multiple evidences for the youth but evolved status of this
star). The weak H$\alpha$ emission and absence of veiling (Lee, Martin,
\& Mathieu 1994) suggest a small or nil accretion rate. Rydgren \& Vrba
(1983a) did not detect IR excess, whereas Weaver \& Jones (1992)
possibly did, indicating that there might still be some material in the
environment; Wolk \& Walter (1996) attribute the IR flux to the
photosphere, and not to optically thin material. There is no evidence
for an associated disk, mass accretion or mass loss (Mathieu et
al. 1989). No CB disk was detected at 2.7~mm (Dutrey et
al. 1996). V826~Tau was not detected at 1100~\micron\ (Skinner, Brown,
\& Walter 1991), so an upper limit to the disk mass based on the
$3\sigma$ upper value of the 1100~\micron\ observations gives an upper
$3\sigma$ limit of 0.03~M$_{\sun}$. The $3\sigma$ upper limit to the
disk's mass, based on 800~\micron\ observations, is 0.007~M$_{\sun}$
(Jensen et al. 1994).


This star also has a photometric period (Rydgren \& Vrba 1983b) of
4.05$\pm$0.2 d. The photometric minima is shifting with time, possibly
due to changes in the spot numbers or position (Reipurth et al.
1990). Stellar spots could introduce polarimetric variations through
non-uniform illumination of the CS material; these variations would then
have a period of $\approx 2.0$ d, half of the photometric period. Mundt
et al. (1983) have deduced from spectroscopic observations an orbital
inclination of $7\fdg2$, compatible with the inclination of 13\arcdeg\
found by Reipurth et al. (1990), a circular orbit and mass ratio of 1.0,
which has been confirmed by Reipurth et al.  (1990) and Lee et
al. (1994). This mass ratio and circular orbit simplify the
interpretation of the polarimetric variations based on our numerical
simulations (Papers~I and II). With such a low inclination, our
numerical models predict variability in position angle ($\Delta \theta
\approx 15\arcdeg$), but not in polarization.

Our polarimetric data for V826~Tau are presented in
Table~\ref{Tab-v826tau} and Figure~\ref{Fig-v826tau}, where the fit made
according to equations~\ref{p3-eq-qfit} and \ref{p3-eq-ufit} is also
presented. Since the IS polarization is very low in the vicinity of
V826~Tau, 0.04\% $\pm$ 0.05\% (see Table~\ref{Tab-averpol} and
Figure~\ref{Fig-Polis_V826_DQ}), the observed polarization, 0.85\% at
67\arcdeg, is probably almost entirely intrinsic, despite the fact that
there is no indication for large amounts of CS material.

M\'enard \& Bastien (1992) presented polarimetry of V826~Tau obtained
with a 4700 \AA\ filter and concluded that the star was possibly
variable. We note that $P$ and $\theta$ are very similar in the blue
(data from 1984, M\'enard \& Bastien 1992) and in the red (this paper),
although the observations were taken years apart.  Our statistical tests
conclude that the star is variable, but neither set of data clearly
shows periodic variations. In particular, we do not see the variability
in position angle that would indicate a low orbital inclination, but
there is a variability in polarization level, although it is not clear
that it is periodic (see Figure~\ref{Fig-v826tau}). To see if spots
could be responsible for the polarimetric variations, half of the
photometric period was used instead of the orbital one, but this does
not reveal any clearer periodic variations. The intrinsic polarization
and polarimetric variability are puzzling in regard of the lack of
evidence for CS material. This indicates that polarimetry might be more
sensitive to the presence of dust than other techniques.

\subsection{UZ~Tau~E/W = HBC~52}
UZ~Tau is a quadruple system; the E and W binaries are $3\farcs78$ apart
(530~AU with an assumed distance of 140~pc), at a position angle of
273\arcdeg\ (Leinert et al. 1993). Our measurements, made with an
$8\farcs2$ aperture hole, include the east and west components, although
the brighter E component contributes more to the measurements.  The west
component, UZ~Tau~W, is a close binary (Simon \& Prato 1995) with a
separation of $\sim 0\farcs35$ (50~AU), at $\sim$0\arcdeg\ position
angle (Leinert et al. 1993; Ghez et al. 1993). It is not surrounded by a
CB disk, and has less material (0.002--0.04~M$_{\sun}$) than the east
component (Jensen, Koerner, \& Mathieu 1996b).  According to Simon \&
Prato (1995), both binaries have $K-N$ values that indicate optically
thick inner disk regions in the IR: UZ~Tau~E has $K-N=2.3$ mag and the W
component has $K-N=2.8$ mag.

UZ~Tau~E is a single-lined spectroscopic binary, with a projected
primary semi-major axis of 0.03~AU (Mathieu, Martin, \& Maguzzu 1996);
its position angle measured eastward from north is $33\pm14$\arcdeg\
(Dutrey et al. 1996). Although no CB disk was detected at 2.7~mm (Dutrey
et al. 1996), Jensen \& Mathieu (1997) argue that the submm continuum
emission must arise in a CB disk. This disk shows no evidence for
central clearing, has a mass of 0.06~M$_{\sun}$ and a radius of 145~AU
(Jensen et al. 1996b), and is a reservoir of material for active
accretion (Mathieu et al. 1996). The SED is well reproduced by a
continuous accreting disk, although the SED and emission feature at
10~\micron\ can also be fit with a partially evacuated hole (Jensen \&
Mathieu 1997). Being a close spectroscopic binary does not remove much
material, as opposed to wider binaries like the W component (Jensen et
al. 1996b).

The structure around UZ~Tau~E has a FWHM of 300~AU, and is oriented at
position angle 19\arcdeg, compatible with the orientation of the binary
($33\pm14$\arcdeg, Dutrey et al. 1996), previous polarization
measurements (Jensen, Mathieu, \& Fuller 1996a), and our first
measurement (see Table~\ref{Tab-autres}). The fact that the position
angles of the disk and polarization are similar indicates that most of
the polarization is intrinsic (otherwise, if it were interstellar, the
polarization angle would not be related to the position angle of the
disk), or that the IS polarization angle is the same as the physical
elongation of the circumbinary disk. The CO emission has a 2:1 aspect
ratio, which suggests that the system is seen more edge-on than pole-on
(Jensen et al. 1996b), which is in agreement with the flattening of the
radio continuum emission and indicates an orbital inclination of about
70\arcdeg\ (Dutrey et al. 1996).

Our 2 measurements, taken a few months apart, are presented in
Table~\ref{Tab-autres}. Polarimetric observations were also made by
Bastien (1982, 1985).  Bastien (1982) presented two measurements made in
the fall 1977, one with a red filter, narrower than ours, and with a
bigger aperture hole of $10\farcs1$. The polarization was 0.45\% at
$2\fdg4$, whereas we measured an average of 0.8\% at 16\arcdeg; the
polarization was then significantly lower in the past but the position
angle could be similar. Two measurements made with a green filter
(5895\AA) in the fall 1977 (Bastien 1982) and winter 1980 (Bastien
1985), and with similar aperture holes ($10\farcs1$ and $12\farcs6$) are
statistically different: 0.47\% at $9\fdg2$ and 1.02\% at $13\fdg7$. Our
2 observations are also statistically different, showing a change of
0.5\% and 45\arcdeg\  in 4 months. All the available data indicate that
there is some variability on time scales of months and years, and that
part of UZ~Tau's polarization is intrinsic. The significant change in
polarization angle possibly indicates modifications of the CS and CB
environments of this quadruple system.

The IS polarization in a region 15\arcdeg\ and $\pm$ 70 pc around
UZ~Tau~E/W is very low, 0.11\% $\pm$ 0.05\% (see Table~\ref{Tab-averpol}
and Figure~\ref{Fig-Polis_V773_Lk_UZ}) pointing to an intrinsic origin for
this binary's polarization, with maybe a small IS component. The long
term variability exposed earlier is compatible with this conclusion.


\subsection{DQ~Tau = HBC~72}
Simon \& Prato (1995) report $K-N=3.5$~mag for both stars of this
15.8-day binary (Stassun et al. 1996), which indicates the presence of an
optically thick inner disk. Stassun et al.  (1996) give a larger NIR
excess ($K-N=4.2$ mag) which, with the emission observed in the mm,
indicates the presence of a massive CB disk, as has also been found by
Jensen \& Mathieu (1997) from submm continuum emission. The SED gives no
indication of an inner hole or gap (Mathieu et al. 1997), but the data
could still be consistent with a small amount of residual warm material
in a central hole. The CB disk would have a mass of
0.002--0.02~M$_{\sun}$ (Mathieu et al. 1997) .

The mass ratio is indistinguishable from unity. Each star is $3\times
10^6$~yr old and has a mass of 0.65~M$_{\sun}$. With an adopted total
mass of $1.3~M_{\sun}$, the inclination is 22\arcdeg\ (Mathieu et
al. 1997). There could be a third star in this system: there is a faint
star $7\farcs3$ from the binary, at position angle $\sim150$\arcdeg;
alternatively, it could be a reddened background star (Mathieu et
al. 1997). There are recurring flare-like events with the same period as
the orbital one, and these occur just before periastron passage, at
least 65\% of the time. With an eccentricity of $e=0.556$, the
periastron separation is 0.060 AU or 13 stellar radii. The flares are
caused by variable accretion (Mathieu et al. 1997). It could be very
interesting to monitor the polarization variations during periastron
passages.

DQ~Tau was observed by Breger (1974), who measured 1.1 $\pm$ 0.3\% at
46\arcdeg, without any filter. Our single observation (see
Table~\ref{Tab-autres}) is 0.6\% at 79\arcdeg. DQ~Tau is located in the
same area as V826~Tau, where the IS polarization is low (see
Table~\ref{Tab-averpol} and Figure~\ref{Fig-Polis_V826_DQ}), but could
contribute to DQ~Tau's polarization.

 
\subsection{NTTS~045251+3016 = HBC~427 = TAP~57}
NTTS~045251+3016 is a long-period spectroscopic binary with a period of
over 2500 d (Mathieu 1994); its projected separation is 4.0~AU (Jensen
et al.  1994).  The IR excess is $K-N=0.6\pm0.2$ (Simon \& Prato 1995)
which might indicate that there is an optically thin inner disk or no
disk at all; Wolk \& Walter (1996) conclude that the minimal excess seen
in $K$ and $L$ could be attributed to cool spots or to the companion,
and not necessarily to optically thin material.  NTTS~045251+3016 was
not detected at 1100~\micron\ (Skinner et al. 1991), so an upper limit
to the disk mass based on the $3\sigma$ upper value of the 1100~\micron\
observations is $0.02~M_{\sun}$.  In a series of 26 $BVR$ photometric
observations obtained during August and September 1992, Grankin (1993)
found variations of the order of 0.15~mag in $V$, but no period could be
found. According to Zakirov, Azimov, \& Grankin (1993), brightness
variations in $V$ had an amplitude of 0.25~mag in 1991/1992, with a
rotational period of 9.32~d; no eclipsing effect could be
found. 

NTTS~045251+3016's low polarization (0.1\% at 107\arcdeg\ - see
Table~\ref{Tab-autres}) may be entirely IS, which would be
compatible with the lack of evidence for CS material. However, its
position angle is different than the one for the IS polarization,
58$\pm$9\arcdeg\ (see Table~\ref{Tab-averpol} and
Figure~\ref{Fig-Polis_NTTS045251}), pointing to an intrinsic origin for part
of the polarization. The polarization was determined to be statistically
constant, although the long period has not been sampled adequately to
reveal orbital variations in the polarization.

\subsection{GW~Ori = HBC~85 = HD~244138}
GW~Ori was discovered by Mathieu et al. (1991) to be a spectroscopic
binary with a period of 241.9~days, a separation $\sim$1~AU and an
eccentricity almost indistinguishable from zero. The center-of-mass
velocity implies a third star with an orbital period of many years or a
$m=1$ perturbation on the disk (Mathieu et al.  1991). It is one of the
brightest CTTS. Based on evolutionary tracks, the primary is massive
(2.5~M$_{\sun}$, Mathieu et al. 1991), and young ($1\times10^6$~yr).

There are large NIR and FIR excesses, with a strong silicate emission
feature, and a dip between 2 and 20~\micron\ that could be due to a gap
from 0.17 to 3.3~AU in the optically thick disk, gap where the secondary
would then be located (Mathieu et al. 1991). Mathieu et al.  (1991) have
considered two models to reproduce the flat SED; the first one is
composed of a circumprimary and a circumbinary disks, and optically thin
dust in the gap between these two disks; the second one is composed of a
circumbinary shell of inner radius 100~AU with a circumbinary
disk. Submillimeter observations are inconsistent with the disk-envelope
model (Mathieu et al. 1993), but indicate GW~Ori is surrounded by a
massive 500~AU circumbinary disk with a model-independent lower mass of
0.3~M$_{\sun}$ (with a range from 0.07 to 0.8~M$_{\sun}$), with an
uncertainty of a factor of 3 due to opacity normalization. The strong
H$\alpha$ emission suggests accretion, but on the other hand, there is
no veiling (Schneeburger, Worden, \& Wilkerson 1979; Basri \& Batalha
1990). A single steady-accretion disk can not reproduce the SED, and a
more luminous CB disk is required (Mathieu et al.  1995). Mathieu et
al. (1995) study again pure-disk and disk-shell models.  The pure-disk
model gives a disk mass of 1.5~M$_{\sun}$, within a factor of three,
which is a significant fraction of the total stellar mass.  The
disk-shell model falls short of reproducing the submillimeter
observations. The large submillimeter emission detected within 500 AU of
the binary is not due to the inner part of a more extended infalling
envelope, but such an envelope could be responsible for the FIR
emission.

A high-resolution interferometric map obtained at 1360~\micron\ reveals
a source size of 1\farcs7 $\times$ 0\farcs8 at position angle 56\arcdeg,
although this result must be considered with caution since the source
size is smaller than the beam size (Mathieu et al. 1995); this angle is
not related with the polarization position angle which we found to be
between 115\arcdeg\ and 130\arcdeg\ (see Table~\ref{Tab-gwori}).

The inclination is reported to be $15\pm1$\arcdeg\ by Bouvier \& Bertout
(1989) although this value is very uncertain. Mathieu et al.  (1991)
argue that if a substantial fraction of the total luminosity of this
system is due to CS material, then this inclination of 15\arcdeg\ is a
lower value, and in fact they found 27\arcdeg. On the other hand,
Shevchenko et al. (1992) found Algol-like fadings of 0.4~mag in $V$ due
to CS material around the secondary star and near phase 0.0, which would
imply an orbital inclination between 80\arcdeg\ and 90\arcdeg.

In 1984 February, Bouvier and Bertout monitored this star to investigate
the presence of periodic photometric variations that could be explained
by the presence of hot or cool spots. The variations observed seem
periodic with a period of 3.2~days, but are of very low amplitude, the
highest amplitude being for the $U$ filter, with an amplitude of $\sim
0.1$ mag. The simple one-spot model could not be satisfactorily applied
to the observations.

The average IS polarization, found by averaging the polarization of 30
stars within 10\arcdeg\ and 200~pc of GW~Ori, is very low in the
vicinity of GW~Ori, 0.04\% $\pm$ 0.04\% (see Table~\ref{Tab-averpol})
but the IS polarization map (see Figure~\ref{Fig-Polis_GWOri}) shows some
polarized stars with aligned vectors. We conclude that its polarization
is mostly intrinsic with maybe an IS component, which is compatible with
the presence of variability (see below). Table~\ref{Tab-gwori} and
Figure~\ref{Fig-gwori} present polarimetric data obtained over a little
more than 3 orbits, along with the fit made according to
equations~\ref{p3-eq-qfit} and \ref{p3-eq-ufit}. The data are clearly
variable, but not periodically. GW~Ori was also observed in polarimetry
by Hough et al. (1981), Bastien (1982, 1985) and M\'enard \& Bastien
(1992). Data taken in the winter 1976 (Bastien 1982) and winter 1980
(Bastien 1985) with the same 5895 \AA\ filter show a drastic difference
that could be intrinsic or due to the different aperture holes used
($14\farcs3$ and $4\farcs3$): 0.21\% at 103\arcdeg\ and 2.64\% at
151\arcdeg. If we compare our observations taken between 1996 and 1999
with an earlier observation made with a red but narrower filter, and a
bigger aperture hole of $14\farcs3$ in winter 1976, we note again a
difference: 0.35\% at 102\arcdeg\ versus 0.61\% at 126\arcdeg\ on
average. There is also a significant rotation of the position angle as a
function of time and wavelength (151\arcdeg\ at 5895 \AA\ in 1980
(Bastien 1985), 95\arcdeg\ at 4700 \AA\ in 1984 (M\'enard \& Bastien
1992), 126\arcdeg\ at 7660 \AA, this work). M\'enard \& Bastien (1992)
classified GW~Ori as a suspected variable; the data presented here
clearly indicate variability. In summary, GW~Ori's polarization and
position angle are variable as a function of wavelength and time, on
time scales of months and years. As mentioned above, this is an
indication in favor of intrinsic polarization, although there also may
be a component of IS polarization.

The fact that it shows variability but no clear periodic variations may
be an indication that important changes in the circumstellar environment
occur over one orbit, or that the star is active. Since the presence of
a massive CB disk is inferred from the spectral energy distribution,
such changes could be possible.

If this binary has a high inclination as claimed by Shevchenko et
al. (1992), rapid changes in the polarization should be observable near
phase 0.0, but our sampling is not sufficient near that phase. On the
other hand, there could be an important variation in the polarization
angle between phases 0.7 and 0.9 (see Figure~\ref{Fig-gwori}). 


\subsection{Par~1540 = HBC~447 = NGC~1977~334}
This 33.73-day binary (Mathieu 1994) does not present photometric
variations over 0.5~mag in $V$; there is no evidence for a significant
IR excess; both components are NTTS (Marshall \& Mathieu 1988). The mass
ratio is 1.32$\pm$0.03; the weak H$\alpha$ emission and absence of
veiling indicate small or nil accretion rate (Lee et al. 1994). Of the
five double-lined PMS binaries studied by Lee (1992) (V826~Tau,
Par~1540, Ori~429, Ori~569, and NTTS~162814-2427), Par~1540 has the
lowest age, $10^5$~yr, and the highest \ion{Li}{1} abundance.

The polarization for Par~1540, 0.83\% at 77\arcdeg\ in the 7660 \AA\
filter (see Table~\ref{Tab-averpol}), is mostly of intrinsic or at least
local (intra-cluster) origin, with a smaller IS component. The
polarization of nearby stars is low and $\approx$0.30\% $\pm$ 0.05\% at
64\arcdeg\ (see the map in Breger 1976, our estimation of the IS
polarization in Table~\ref{Tab-averpol}, and a map of IS polarization in
Figure~\ref{Fig-Polis_Parenago}). Par~1540's extinction is $E(B-V)=0.35$
(Marshall \& Mathieu 1988), while the reddening between Earth and the
Orion Nebula is only $E(B-V)=0.05$ (Breger, Gehrz, \& Hackwell
1981). The polarimetric variability (see below) also indicates the
presence of intrinsic polarization.  Par~1540's position angle is
77\arcdeg, typical of the less polarized stars studied by Breger (1976),
and similar to the average (64\arcdeg) of stars within its neighborhood
and with similar distance. This indicates the presence of an IS
polarization component.

Our data are presented in Table~\ref{Tab-p1540},
Figures~\ref{Fig-p1540a} and \ref{Fig-p1540b}. Breger (1976) measured a
polarization of $1.11 \pm 0.10$\% at $85\arcdeg$, without any filter
(giving an effective wavelength between $B$ and $V$). This value is
above the average of our measurements, but this could be due to the
wavelength dependence of the polarization. Data obtained with the
polarimeter STERENN at the Pic-du-Midi Observatory (France) with a $V$
filter and a 10\arcdeg\ aperture hole on 14 January 1994 gives a
polarization of 0.992$\pm$0.129\% at 83\arcdeg\ (F. M\'enard 2001, private
communication), very similar to Breger's measurement and possibly
indicating long-term stability, although our data is clearly variable,
in polarization and position angle, on time scales of months and years.

In Figure~\ref{Fig-p1540a}, one atypical point is not shown; many binary
PMS stars, like Par~1540, showed atypical values of the polarization
that were removed from the data set before fitting the
data. Nonetheless, the data is very noisy, and hard to fit according to
equations~\ref{p3-eq-qfit} and \ref{p3-eq-ufit}. Statistical tests
conclude that the polarization is variable. In Figure~\ref{Fig-p1540b},
the data were binned by dividing the phase into 10 bins and averaging
all the polarization observations in each bin.  This makes small
amplitude periodic variations stand out more clearly, and the quality of
the fit is increased.  The LNP shows two peaks at 21.3 and 16.2 d, with
only slightly better than 50\% chance of not being a signature of random
Gaussian noise; the former period is also found by the PDM while the
latter could reflect the double-periodic low amplitude variations that
appear in the binned data (see Figure~\ref{Fig-p1540b}).


\subsection{Par~2486 = NGC~1977~1060 = BD~$-$05~1340}
The mean polarization of that 5.1882-day binary (Mathieu 1994), 0.14\%
at 63\arcdeg\ (see Table~\ref{Tab-p2486} and Figure~\ref{Fig-p2486}), is
mostly IS with also an intrinsic component. Located near
Par~1540 (see map in Figure~\ref{Fig-Polis_Parenago}), it is in a
region of low IS polarization. The average IS position angle is
73\arcdeg, close to the observed one. Statistical tests indicate
possible variability in polarization and position angle, on time scales
of months and years. More polarimetric observations would help confirm
the variability

\subsection{Ori~429 = ORINTT~429}
This 7.46-day binary (Mathieu 1994) has a mass ratio is
1.01$\pm$0.07. The weak H$\alpha$ emission and absence of veiling
indicate a small or nil accretion rate (Lee et al.  1994). It is located
in the same region as the two previous stars (see also map in
Figure~\ref{Fig-Polis_ORI429_569}). Data are presented in
Table~\ref{Tab-ori429} and Figure~\ref{Fig-ori429}.  The mean
polarization at 7660 \AA, 0.2\% at 72\arcdeg, is mostly IS, as it is
similar to the estimated IS polarization. Statistical tests do not
indicate any variability.

\subsection{Par~2494 = HBC~487 = NGC~1977~1069}
Data for this 19.4815-day binary (Mathieu 1994) are presented in
Table~\ref{Tab-p2494}, Figures~\ref{Fig-p2494a} and
\ref{Fig-p2494b}. This star is located in the same region as the three
previous objects, where the IS reddening and polarization are low (see
Table~\ref{Tab-averpol} and Figure~\ref{Fig-Polis_Parenago}). Since its
position angle, $46\arcdeg$, does not correspond to the most common
value for the less polarized stars measured by Breger (1976), nor to the
IS position angle of 78\arcdeg\ for stars with similar
distances, part of the polarization is intrinsic, but IS polarization is
also present.

The mean polarization at 7660 \AA, 0.16\% at 46\arcdeg, was determined
to be statistically constant. Breger (1976) measured a polarization of
$0.54 \pm 0.11$\% at $69\arcdeg$, without any filter, which is above the
mean of our measurements, but could be due to the wavelength dependence
of the polarization. Data obtained with the polarimeter STERENN at the
Pic-du-Midi Observatory (France) with a $V$ filter and a 10\arcdeg\
aperture hole on 29 January 1992 gives a polarization of 0.41$\pm$0.14\%
at $52\fdg5$ (F. M\'enard 2001, private communication), similar to Breger's
measurement and pointing to stability.  Although all the available data
point to a stable polarization, the binned data presented in
Figure~\ref{Fig-p2494b} show periodic variations, in position angle at
least, and it is possible to get a reasonable fit when using
equations~\ref{p3-eq-qfit} and \ref{p3-eq-ufit}. Since the statistical
tests we used to check for the presence of variability do not take into
account low amplitude but systematic variations such as those seen in
the position angle for Par~2494, there could be periodic variations. The
PDM and LNP find similar period of $\sim 38.5$ d for Par~2494, with a
70\% chance that the data are not random Gaussian noise; this period is
about twice the orbital period.

\subsection{Ori~569 = ORINTT~569}
This 4.25-day binary (Mathieu 1994) has a mass ratio of
1.00$\pm$0.02. The weak H$\alpha$ emission and absence of veiling
indicate small or nil accretion rate (Lee et al.  1994). Lee (1992)
discovered a third star in this system. Ori~569 is at about 3\arcdeg\
from Ori~429 (see map in Figure~\ref{Fig-Polis_ORI429_569}). Data are
presented in Table~\ref{Tab-autres} and Figure~\ref{Fig-ori569}.  The
mean polarization at 7660 \AA\ is 0.18\% at 76\arcdeg, very similar to
the estimated IS polarization around this star (0.21\% $\pm$ 0.04 at
90\arcdeg). Statistical tests indicate possible variability, so we
conclude that the observed polarization is a sum of intrinsic and IS
polarizations.

\subsection{W~134 = Walker~134 = NGC~2264~134 = VSB~92}
Both components of this 6.3532-day binary are G stars showing strong
\ion{Li}{1} 6707\AA\ absorption features; the total mass of the system
is ${\rm M} \sin^3 i = 3.16 {\rm M}_{\sun}$ with a mass ratio of 1.04
(Padgett \& Stapelfeldt 1994). It has a NIR excess more typical of CTTS,
that can not be entirely attributed to dark spots, and that indicates
some warm dust resides within 0.3~AU of the binary; on the other hand,
its weak emission lines make it a WTTS (Padgett \& Stapelfeldt 1994). It
has not been detected longward of 12~\micron\ so it probably does not
have a CB disk (Jensen \& Mathieu 1997).

When using an approximate $v \sin i$, the orbital period and the derived
stellar radii, and the assumption that the system is tidally locked, the
inclination is then $46\arcdeg \pm ^{21}_{15}$ (Padgett \& Stapelfeldt
1994). Theoretical masses give an orbital inclination of $63\arcdeg\pm
4$, which could be sufficient for grazing eclipses, given the stellar
radii derived, but photometric monitorings during zero velocity
separation events did not show any decrease in brightness (Padgett \&
Stapelfeldt 1994). Young (1978) reports $E(B-V)=0.08$~mag for W~134 and
notes that intra-cluster clouds not randomly distributed cause
differential reddening, whereas Padgett \& Stapelfeldt (1994)
recalculated $E(B-V)=0.2$~mag. Koch, Perry, \& Kilambi (1994) report
0.35~mag variability in $V$ and $R$, but no phase-locked variability
could be found, and if present, could be hidden by non-periodic
variations.

Polarization observations of 34 stars belonging to NGC~2264 (Corporon et
al. in preparation) have an average position angle of 16\arcdeg, whereas
the average IS position angle of stars with similar distance
to W~134 is 177\arcdeg, different that W~134's value of 32\arcdeg. A map
of the polarization of neighboring stars is presented in
Figure~\ref{Fig-Polis_W134_VSB126}; the IS polarization vectors are
mostly aligned to each other, indicating the presence of IS polarization
(average of 0.87\% $\pm$ 0.12\%). Since W~134 shows polarimetric
variations (see below), we conclude that its polarization is intrinsic
with an IS component.

Data are presented in Table~\ref{Tab-w134} and Figure~\ref{Fig-w134},
where the fit according to equations~\ref{p3-eq-qfit} and
\ref{p3-eq-ufit} is also shown. The average polarization, 0.22\% at
32\arcdeg, is statistically variable. In Figure~\ref{Fig-w134}, two
atypical position angle values taken more than a year apart can be seen,
near phase 0.55; such rapid changes are sometimes seen for eclipsing
binaries (for example in the eclipsing binary EK~Cep, Manset \& Bastien
in preparation). The grazing eclipses, predicted by Padgett \&
Stapelfeldt (1994) but not seen in photometry, should occur near phases
0.25 and 0.75. The rapid change in position angle is seen near phase
0.55 and not at the predicted phases for the eclipses; nonetheless, more
polarimetric data should be taken near phase 0.55 to investigate the
cause of the atypical polarimetric observations.


\subsection{VSB~126 = NGC~2264~169}
Data for this 12.924-day binary (Mathieu 1994) are presented in
Table~\ref{Tab-vsb126} and Figure~\ref{Fig-vsb126}. The mean
polarization at 7660 \AA, 0.16\% at 66\arcdeg, is statistically
constant. VSB~126 is within 0.2\arcdeg\ of W~134 (see map in
Figure~\ref{Fig-Polis_W134_VSB126}), and therefore is also affected by an
 IS polarization component at 177\arcdeg. Its low and constant polarization,
with an average of 0.16\% at 66\arcdeg, is intrinsic with an IS
component.

\section{Orbital inclination \label{sect-incl}}
One of the goals of these polarimetric observations is to determine the
orbital inclinations of the selected PMS binaries, by using the BME
formalism. This formalism can still be used if the orbits are
non-circular and the scatterers are spherical grains, within the limits
presented in Papers~I and II, and even if the BME formalism, with its
first and second harmonics only, does not reproduce exactly the
variations of eccentric systems.

Noise with a standard deviation greater than 10\% of the amplitude of
the polarimetric variations will prevent the BME formalism from finding
a reasonable estimate of the true inclination (Paper~I). Other studies
have also shown that the quality of the data (number of data points,
observational errors, amplitude of the polarimetric variations) can
strongly influence the results found by the BME formalism.

Aspin, Simmons, \& Brown (1981) have studied what standard deviation
$\sigma_{\rm nec}(i)$ is necessary to determine the inclination $i$ to
$\approx \pm 5 \arcdeg$, with a 90\% confidence level. They give an
approximate relation for the data quality $DQ$:\\
\begin{equation}
DQ = \frac{\sigma_{\rm o}}{A_{\rm o} \sqrt{N_{\rm o}}} =
\frac{\sigma_{\rm nec}(i)}{A(i) \sqrt{N}}, 
\end{equation}
where 
\begin{equation}
A = \frac{|Q_{\rm max} - Q_{\rm min}| + |U_{\rm max} - U_{\rm min}|}{4},
\label{EQ-A}
\end{equation}
$\sigma_{\rm o}$ is the observational error on the polarization, $A_{\rm
o}$ is the observed polarimetric variability calculated with
Equation~\ref{EQ-A}, $N_{\rm o}$ is the number of observations, and
$N=40$ (the number of bins in their simulations). A set of very good
quality observations will have a low value of $DQ$. We present in
Table~\ref{Tab-noiseBME}, Column 2, $DQ$ values for the binaries studied
here. After the quantity $\sigma_{\rm nec}(i) / A(i)$ is calculated,
Table 1 in Aspin et al. (1981) gives the lowest possible inclination
that can be determined from the observations with a $\pm\ 5 \arcdeg$
accuracy at a significance of 10\% (meaning that the true inclination
has a probability of 90\% to be in within 5\arcdeg\ of the value
returned by the BME formalism). If we apply this method to our sets of
data, we find that the quality of our data do not allow us to find $i$
to $\approx \pm 5 \arcdeg$, with a 90\% confidence level, for any of our
binaries.

Wolinski \& Dolan (1994) have also studied the confidence intervals for
orbital parameters determined polarimetrically. They made Monte Carlo
simulations of noisy polarimetric observations, for a specific geometry
not suitable for the stars studied in this present paper, but their
results are nonetheless instructive. Confidence intervals for $i$ are
given graphically as a function of a ``figure of merit'' $\gamma$:
\begin{equation}
\gamma = \left( \frac{A}{\sigma_p} \right) ^2 
\left( \frac{N}{2} \right),
\end{equation}
where $\sigma_p$ is the standard deviation of the noise that was added
to the data and $N$ is the number of observations. We have calculated
and present in Column 3 of Table~\ref{Tab-noiseBME} the figures of
merit $\gamma$ for some PMS binaries, by using the observational error
$\sigma(P)$ instead of the $\sigma_p$ used by Wolinski \& Dolan. It is
again seen that the quality of the data is not very good, mostly because
the amplitude $A$ is rather low (between 0.02 and 0.10\% in general).

Finally, following our own studies of the effects of noise on the BME
results (Paper~I), we have calculated the noise for the Stokes
parameters $Q$ and $U$, by using the variance of the fit and the
amplitude of the polarimetric variations; these amplitudes are computed
from the maximum and minimum values of the observations, and not those
of the fit. These calculations are presented in Columns 4 and 5 of
Table~\ref{Tab-noiseBME}, where levels of noise below 10\% are
rare. Once again, this analysis shows that the polarimetric observations
of PMS stars are not of ``very good quality'', not because of
instrumental or observational problems, but because non-periodic
stochastic polarization variations and low amplitude periodic variations
make the data rather noisy, hiding the periodic polarimetric variations
that most probably exist.

One of the stars for which the polarimetric variations are very clear
and might be of good enough quality for the BME formalism to work is
AK~Sco (Manset \& Bastien, in preparation); this star is the only one
for which the data were obtained within a few (12) consecutive
nights. All the other stars were observed over 3-5 years. For these, non
periodic polarimetric events, or changes in the circumstellar
environment from epoch to epoch, mask the periodic variations and
introduce too much noise. Future observations should be obtained at a
site offering many consecutive clear nights to cover in one run the
whole orbital period.

Assuming the BME formalism can be used to analyze the polarimetric
variations of binary PMS stars, we have added in
Table~\ref{Tab-noiseBME} the results of the BME analysis for the orbital
inclination. Most of the inclinations are near 90\arcdeg, which cannot
be a real result. This is compatible with the above discussion on the
effects of the stochastic noise on the inclination analysis, and does
not necessarily mean that the BME analysis can not be used for these
systems.

\section{Orientation of the orbital plane and moments of the
distribution of the scatterers} 
In addition to the orbital inclination, the BME formalism returns
$\Omega$, the orientation of the orbital plane with respect to the plane
of the sky, and moments of the distribution of the scatterers, which are
used to measure the asymmetry with respect to the orbital plane ($\tau_0
G$), and the degree of concentration towards the orbital plane ($\tau_0
H$). It is generally expected that the distribution will be symmetric
and concentrated in the orbital plane, so $\tau_0 H > \tau_0 G$. In
Table~\ref{Tab-omegagammas}, we present the values of $\Omega$, $\tau_0
H$, $\tau_0 G$, and the ratio $\tau_0 H / \tau_0 G$. If the circumbinary
disks of these binaries can be imaged (with interferometric or adaptive
optics techniques), their orientation should be similar to $\Omega$,
although the orbits are not necessarily coplanar with the disks or
envelopes. The values for $\tau_0 G$ and $\tau_0 H$ are similar to those
we have found in numerical simulations (Paper~I), and to the observed
values for other types of binaries (Bastien 1988; Koch et al. 1994). In
particular, $\tau_0 H > \tau_0 G$ as expected, with ratios approximately
from 1.0 to 4.0.

These parameters are also calculated using the coefficients of the fits
and the same assumptions (the scatterers are electrons, the orbits are
circular). However, single-periodic variations that are not present when
the scatterers are electrons and the orbits are circular, do appear in
our simulations when there are dust grains instead of electrons,
variable optical depth effects are considered, or the orbits are
eccentric (Paper~I and II). Therefore, the values of the parameters
calculated might not reflect an asymmetric configuration ($\tau_0 G$) or
a concentration toward the orbital plane ($\tau_0 H$). For example, we
have found, using the code presented in Papers~I and II, that even
though $\tau_0 G$ should be null for a perfectly symmetric
configuration, it will not be if we have dust grains or consider
variable optical depth effects.


\section{Discussion}
We have presented polarimetric observations for 14 spectroscopic PMS
binaries located in the Taurus, Auriga, and Orion SFRs. The majority of
the PMS binaries observed have a detectable linear polarization; only
LkCa~3, NTTS~045251+3016, Ori~569, and VSB~126 do not present
polarization levels above the $3\sigma$ limit. For most binaries, IS
polarization is also an important component of the observations. After
an estimation of this IS polarization is removed, a few (5 out of 14) of
the binaries present clear indications of intrinsic polarizations
$\gtrsim 0.5\%$ and significantly above the IS polarization: V826~Tau,
UZ~Tau~E/W, DQ~Tau, GW~Ori, and Par~1540. Overall, 7 of our 14 binaries,
including all the clearly identified CTTS binaries (UZ~Tau~E, DQ~Tau,
and GW~Ori; Mathieu et al. 1997), show intrinsic polarizations above
0.5\%. Interestingly, then, even the WTTS and NTTS, for which other
types of observations are not indicative of significant amounts of
circumstellar material, can have detectable levels of polarization.

As has been found for single PMS stars, the polarization of PMS binaries
is variable. All binaries are statistically variable or suspected
variable, except LkCa~3, NTTS~045251+3016, Par~2494, Ori~429, and
VSB~126. Despite the results of those tests, graphs of polarimetric
data as a function of orbital phase show that LkCa~3 could be variable
in position angle, and Par~2494 is clearly variable in $\theta$. Not
enough data are available for DQ~Tau to determine its variability. When
combining data from literature and our two observations for UZ~Tau~E/W,
variability is present. Therefore, we find that 54\% of the PMS binaries
are variable (7 of out 13), or 77\% (10 out of 13) variable or possibly
so; these results are compatible with those of Bastien (1988) and
M\'enard \& Bastien (1992). All the known CTTS binaries in our sample
(except maybe DQ~Tau for which we only have one measurement) present
polarimetric variations. Many NTTS or WTTS also show polarimetric
variations. Therefore, around these stars, although it is not thought
that there is much CS material, there is enough to produce polarimetric
variations, as it can be shown that very little mass in the form of dust
is needed to produce detectable levels of polarization.

Some stars have shown, once or even a few times, atypical values of
polarization and/or position angle that are well below or above the rest
of the data (Par~1540, Par~2494, and also MWC~1080, Paper~III). We
believe these are real observations of events that strongly affected the
stars and/or their environment. Single PMS stars are known sometimes to
be strongly variable, so this is not a surprise. For the stars with a
sufficient amount of data, additional observations at similar phases
indicate that these atypical points are not related to the normal
periodic behavior. Photometry or spectroscopy obtained at the same times
as those of the atypical observations could help reveal the cause.

Only 6 binaries have enough observations to investigate the presence of
periodic polarimetric variations: LkCa~3, V826~Tau, GW~Ori, Par~1540,
Par~2494, and W~134. Statistical tests conclude that LkCa~3's and
Par~2494's polarization is constant, but low amplitude periodic
variations are nonetheless present. The position angle for LkCa~3's
polarization outlines a sinusoidal wave between phases 0.2 and 0.65;
since the data were obtained over 22 months, representing more than 600
orbits, it cannot be attributed to coincidence, but to a real
variation. Although Par~2494's observations are rather noisy, binned
data reveal low amplitude periodic variations in position
angle. Therefore, low amplitude variations, especially if masked by
noise, can be missed by statistical tests for variability, because these
only consider the whole of the observations without their known order
(as a function of phase).

A third binary, Par~1540, shows periodic variations, although they also
are of low amplitude. Period analysis for Par~1540 and Par~2494, with
PDM and LNP, confirm that the periodicity is due to the orbital motion,
although not with a very high significance. Non-periodic or
pseudo-periodic polarimetric variations could explain why it is
difficult to see periodic variations and confirm them with period
analysis methods. Those non-periodic variations may be caused by
appearance or disappearance of hot or cool spots, flares, or major
changes in the distribution or density of matter in the CS or CB
environment. To avoid or at least decrease the effects of those events,
observations should be carried on consecutive nights before any major
stochastic change in polarization occurs, and until the orbital period
has been sufficiently covered. Those three PMS binaries which exhibit low
amplitude polarimetric variations are WTTS; this indicates that there is
still enough dust in their environment to produce polarization and
periodic variations.

A fourth interesting case is W~134, which may be an eclipsing system:
two atypical position angles, taken more than a year apart, are seen near
phase 0.55; such rapid changes are sometimes seen for eclipsing binaries
(for example in the eclipsing binary EK~Cep, Manset \& Bastien in
preparation; see also St-Louis et al. 1993). The rapid change in
position angle is not at the predicted phases for the eclipses;
nonetheless, more polarimetric data should be taken near phase 0.55 to
investigate the cause of the atypical polarimetric observations.

One of the goals of these observations was to find the orbital
inclinations. Unfortunately, non-periodic or pseudo-periodic variations
sometimes mask the truly periodic variations by introducing noise. This
noise, as measured by 3 different methods, is too high for the BME
formalism to find reasonable estimates of the orbital inclination. Three
factors contribute to this difficulty. First, dust grains are the main
scatterers in these systems, and it has been shown that dust grain
produce polarimetric variations of smaller amplitude than electrons
(Paper~II). Second, the disk around these short-period binaries are
probably CB rather than CS ones, and CB disks produce variations of
smaller amplitudes than CS disks (Paper~II). Finally, non-periodic
events introduce noise that masks the already small amplitude
variations. This last problem might only be improved by taking data on
shorter periods of time.

Other parameters are returned by the BME formalism: $\Omega$, the
orientation of the orbital plane with respect to the plane of the sky,
and moments of the distribution of the scatterers, used to measure the
asymmetry with respect to the orbital plane ($\tau_0 G$), and the degree
of concentration towards the orbital plane ($\tau_0 H$). Although the
assumptions used in the BME formalism (scattering on electrons, circular
orbits) are not met in the PMS binaries studied here and we have shown
that even in simple cases the values are incorrect, the values returned
for PMS binaries are of the same order of magnitude as values for other
types of stars and as for our simulations.

A similar detailed analysis will be presented for PMS binaries in the
Scorpion and Ophiucus SFRs in a coming paper.

\acknowledgments 
N. M. thanks the directors of the Mont M\'egantic Observatory for
granting generous time over many years. The technical support from the
technicians of the observatory, B. Malenfant, G. Turcotte, and F. Urbain
is duly acknowledged. We thank F. M\'enard for providing unpublished
data for Par~1540 and Par~2494. N. M. thanks the Conseil de Recherche en
Sciences Naturelles et G\'enie of Canada, the Fonds pour la Formation de
Chercheurs et l'Aide \`a la Recherche of the province of Qu\'ebec, the
Facult\'e des Etudes Sup\'erieures and the D\'epartement de physique of
Universit\'e de Montr\'eal for scholarships, and P. B. for financial
support. We thank the Conseil de Recherche en Sciences Naturelles et
G\'enie of Canada for supporting this research. N. M. is Guest User,
Canadian Astronomy Data Centre, which is operated by the National
Research Council, Herzberg Institute of Astrophysics, Dominion
Astrophysical Observatory.


\newpage

\scalebox{0.9}{\includegraphics{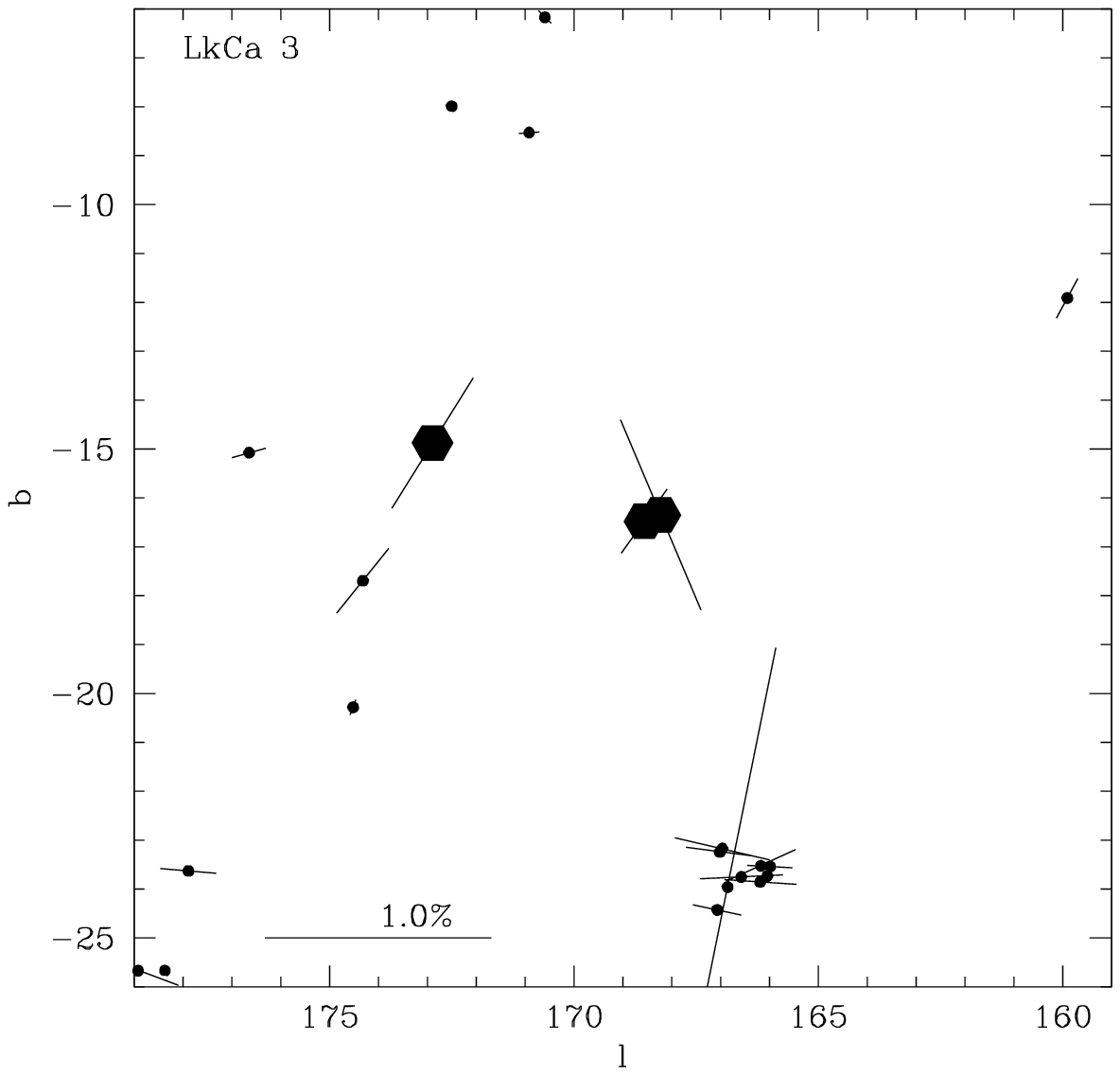}}
\figcaption[Manset4.fig01.ps]{Map of the interstellar polarization in
the vicinity of V773~Tau (hexagonal symbol at right), LkCa~3 (at the
center of the map), and UZ~Tau (left). The stars selected to calculate
the IS polarization are within 70~pc of those
targets. \label{Fig-Polis_V773_Lk_UZ}}

\newpage
\scalebox{0.75}{\includegraphics{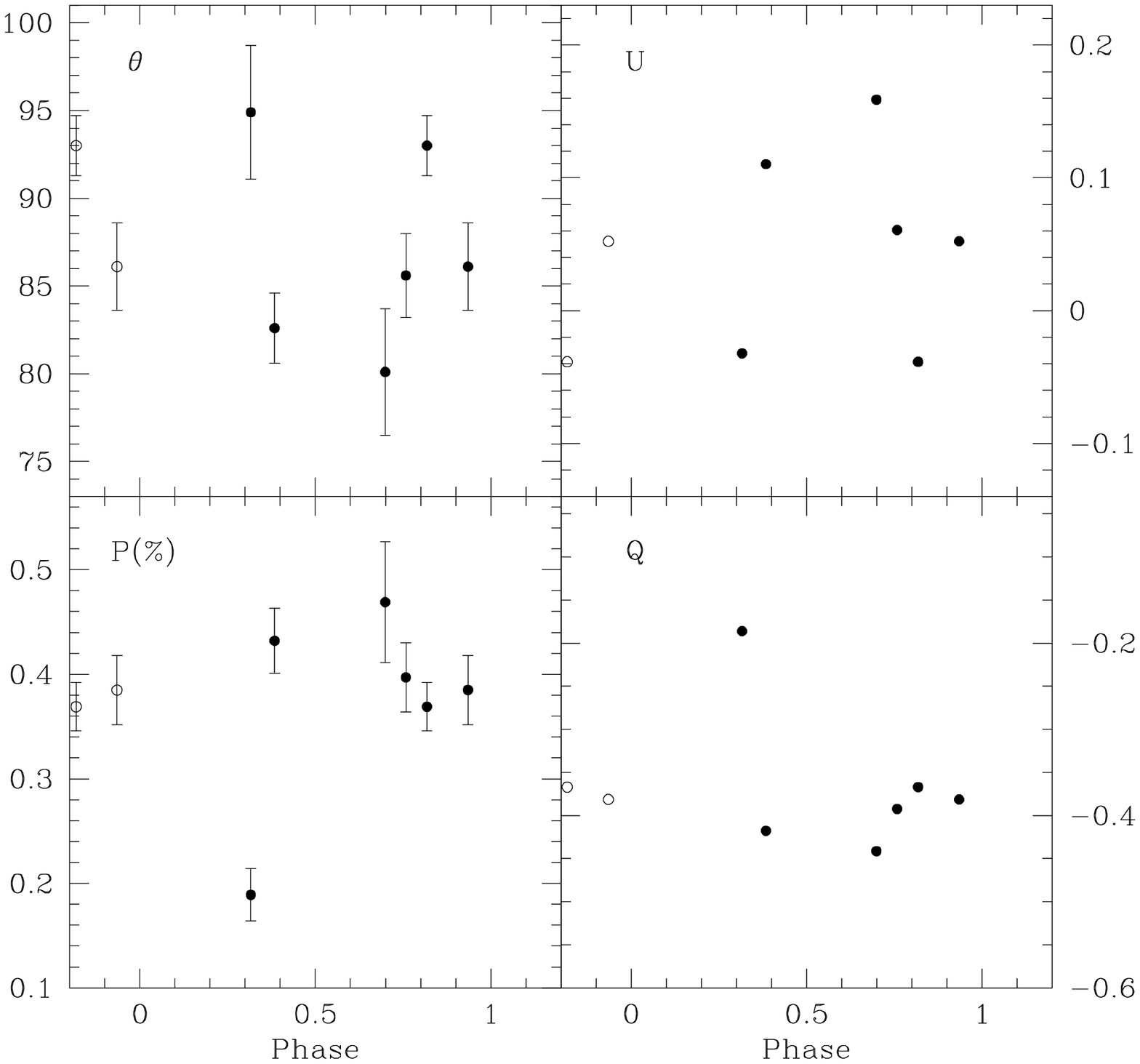}}
\figcaption[Manset4.fig02.ps]{Polarimetric observations of
V773~Tau. This star is polarimetrically variable. \label{Fig-v773tau}}

\newpage
\scalebox{0.75}{\includegraphics{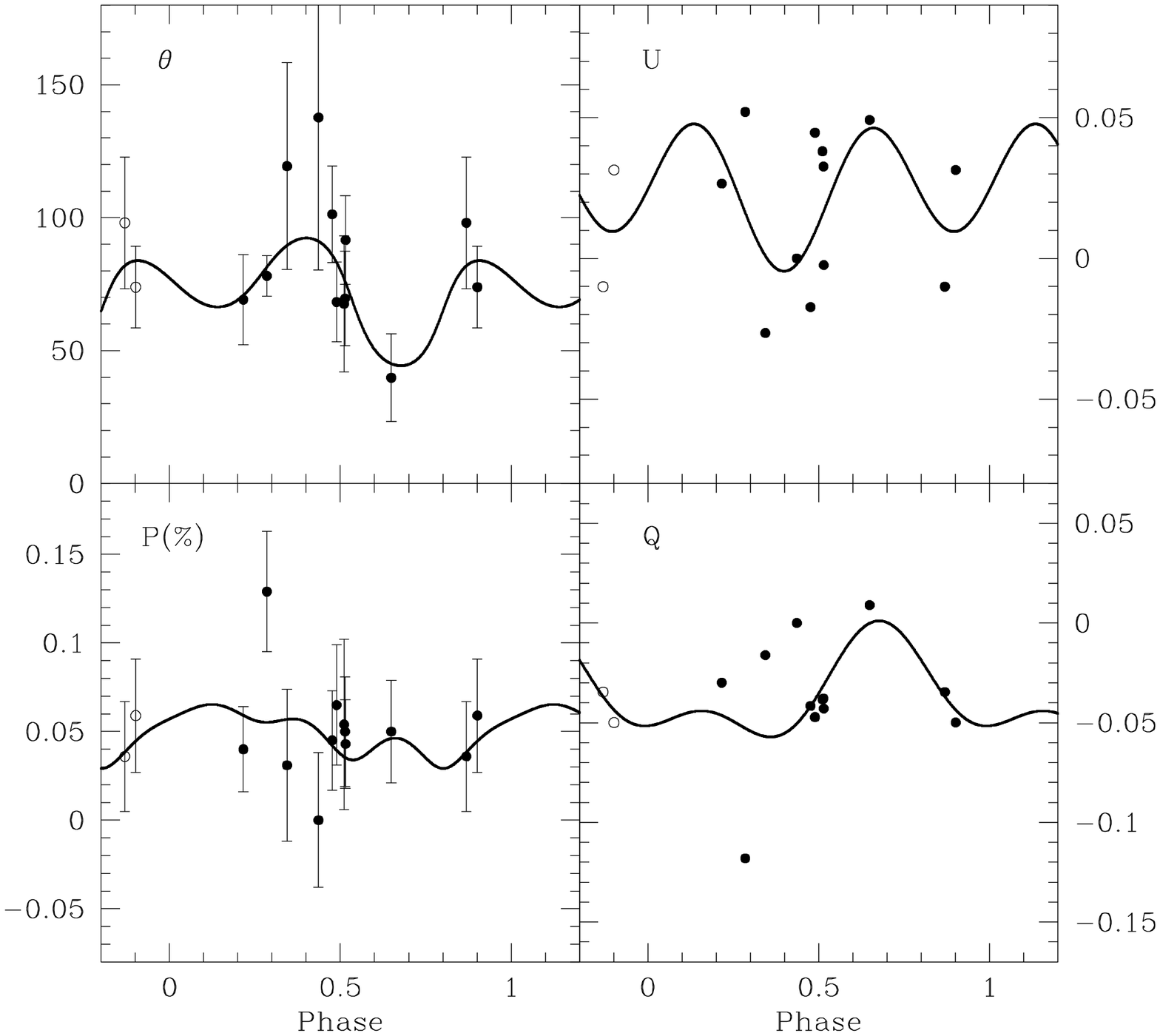}}
\figcaption[Manset4.fig03.ps]{Polarimetric observations of
LkCa~3. Statistical tests determined its polarization is constant,
although there seems to be a sinusoidal trend in polarization angle
between phases 0.2 and 0.65.\label{Fig-lkca3}} 

\newpage
\scalebox{0.75}{\includegraphics{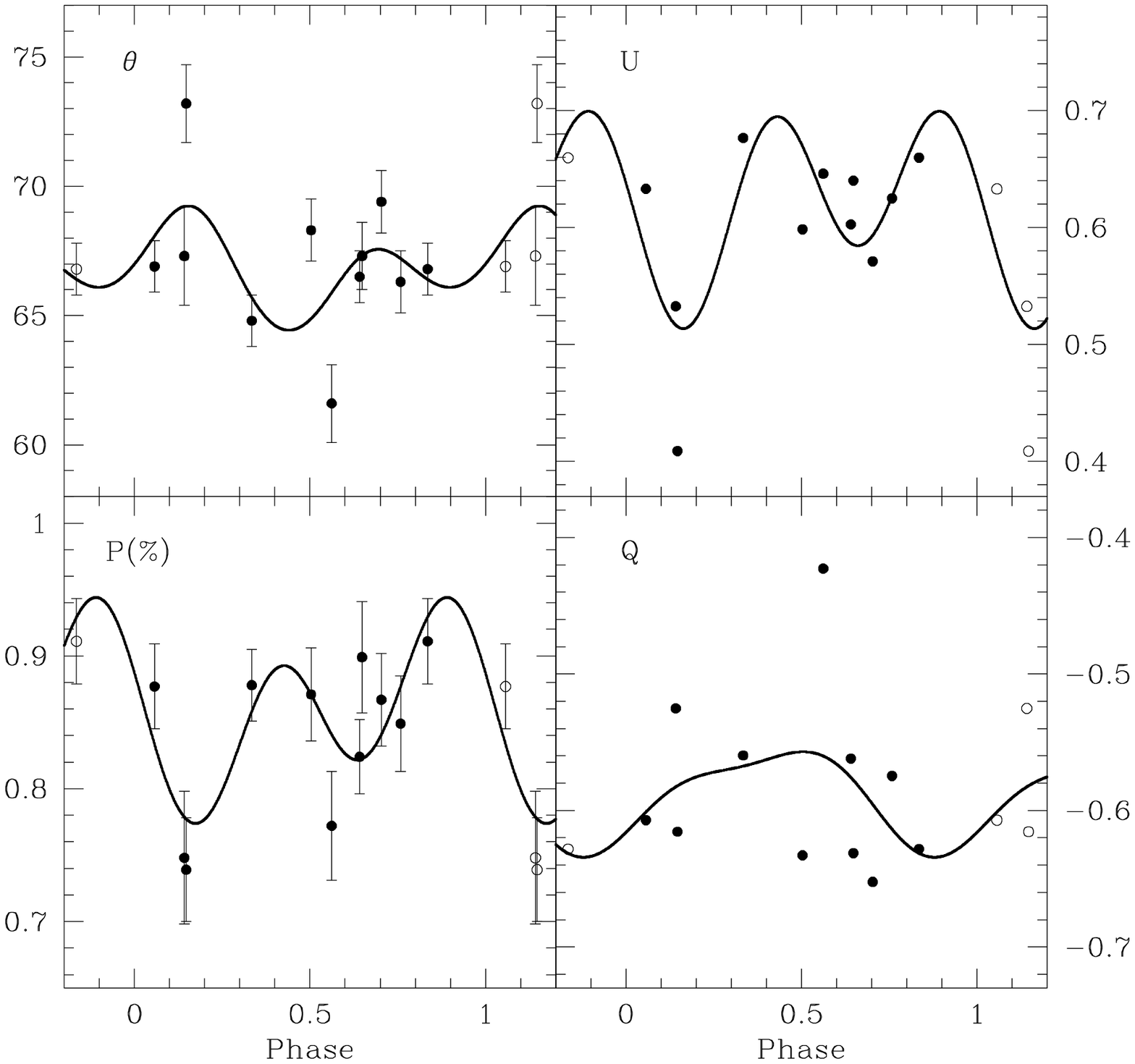}}
\figcaption[Manset4.fig03.ps]{Polarimetric observations of V826~Tau, a
polarimetrically variable binary. \label{Fig-v826tau}}

\newpage
\scalebox{0.9}{\includegraphics{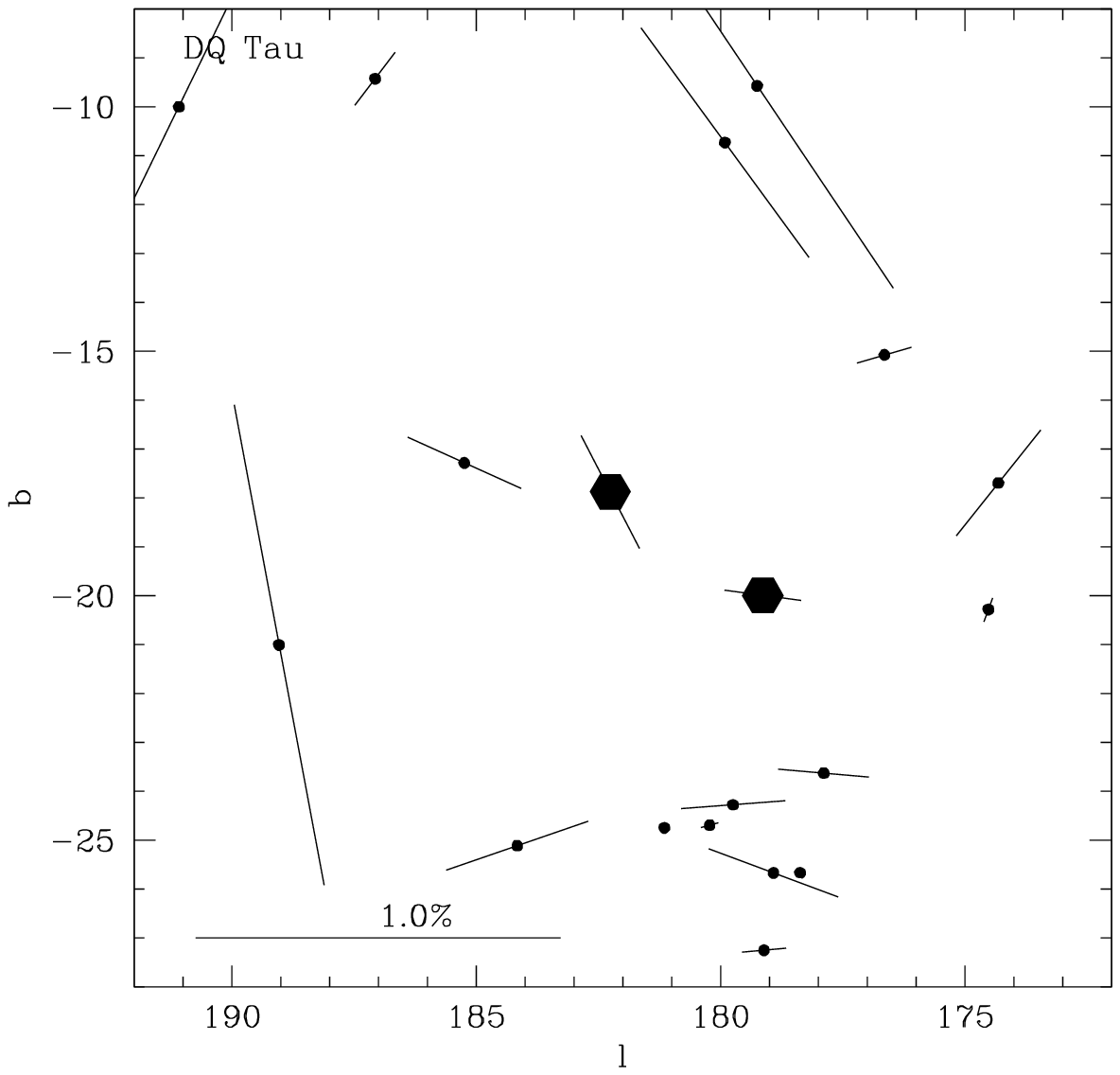}}
\figcaption[Manset4.fig03.ps]{Map of the interstellar polarization in
the vicinity of V826~Tau (hexagonal symbol at right) and DQ~Tau (at the
center of the map). The stars selected to calculate the IS polarization
are within 80~pc of those targets. \label{Fig-Polis_V826_DQ}}

\newpage
\scalebox{0.9}{\includegraphics{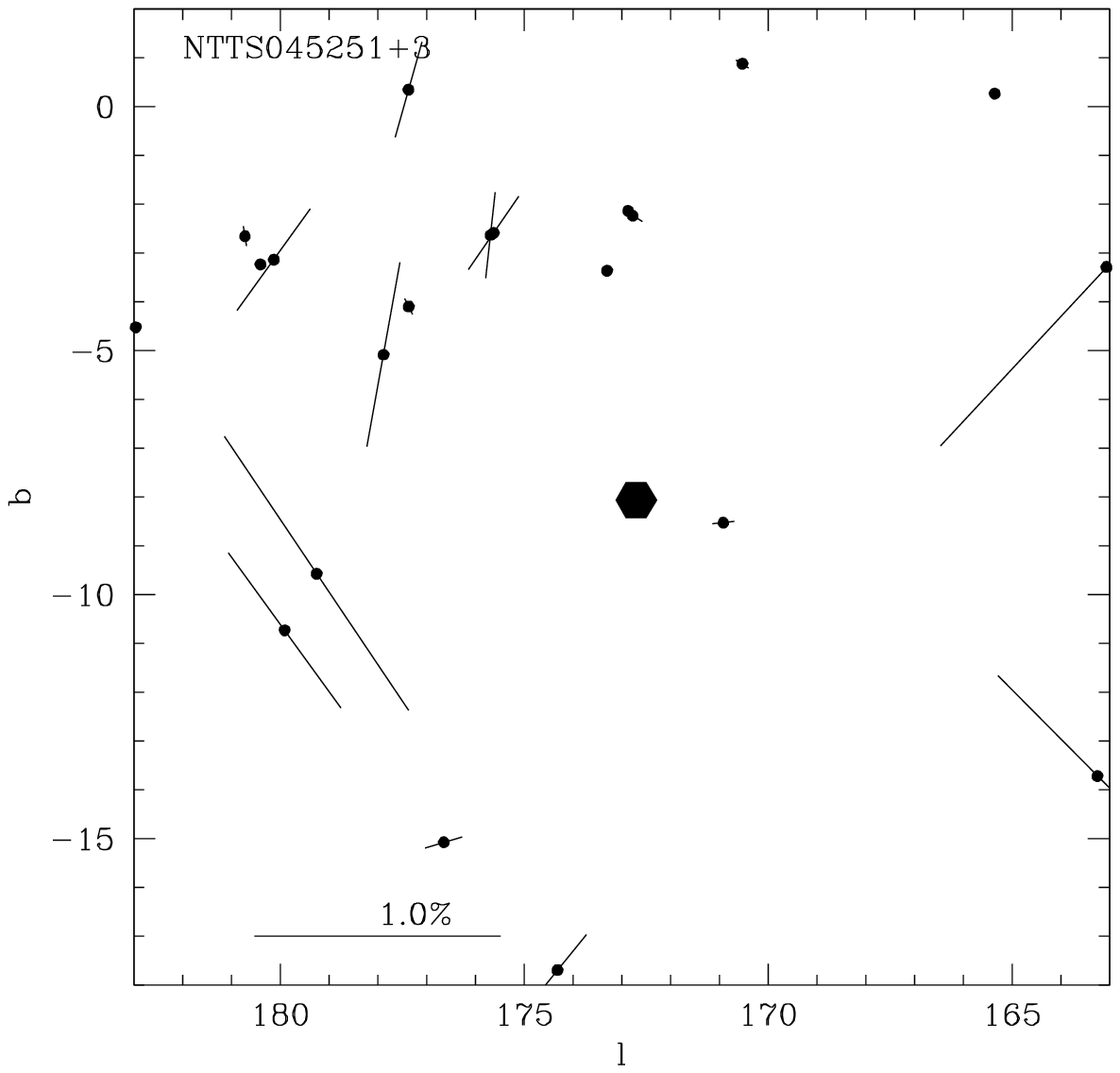}}
\figcaption[Manset4.fig03.ps]{Map of the interstellar polarization in
the vicinity of NTTS~045251+3016. The stars selected to calculate the IS
polarization are within 80~pc of this
target. \label{Fig-Polis_NTTS045251}}

\newpage
\scalebox{0.9}{\includegraphics{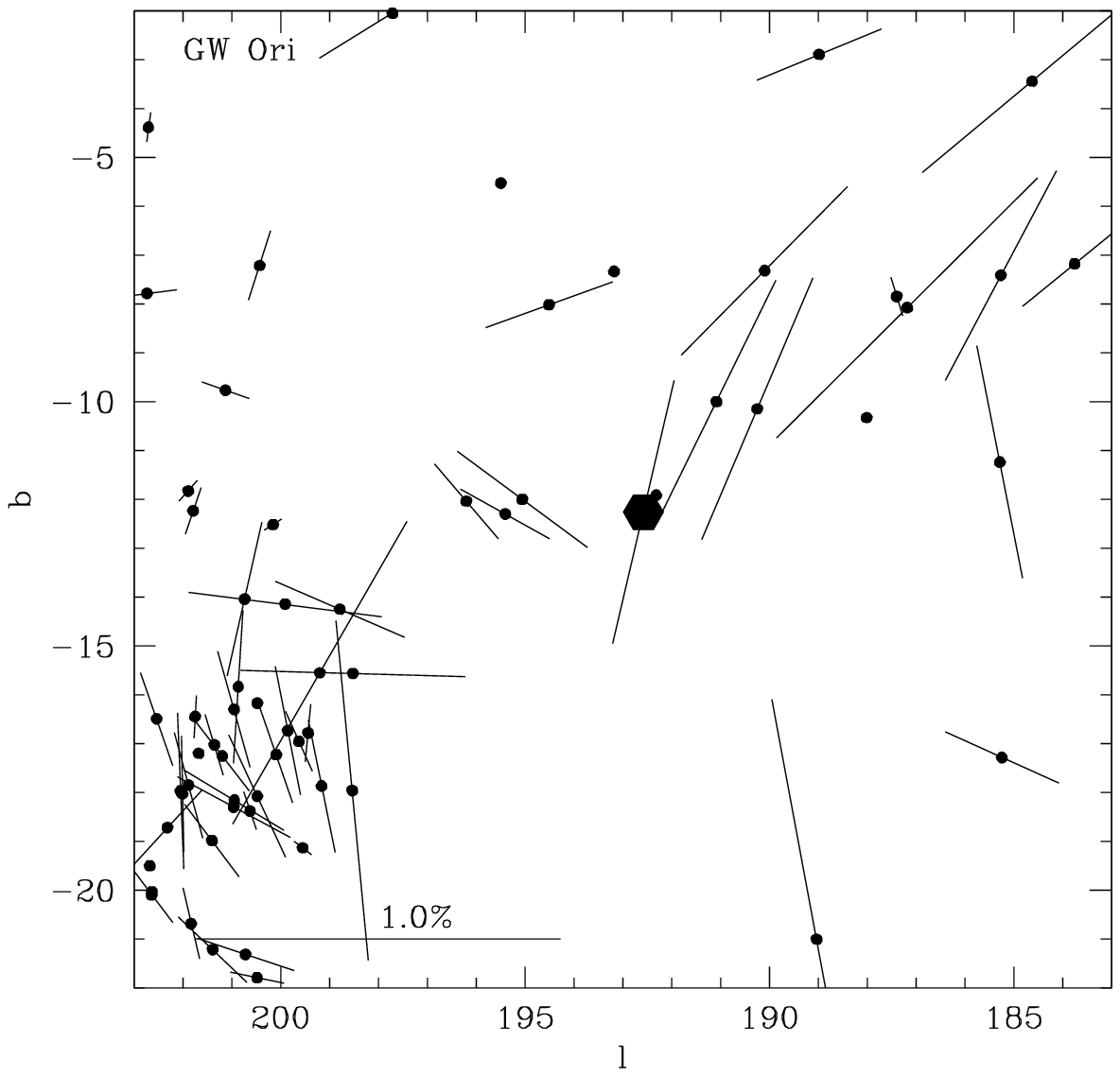}}
\figcaption[Manset4.fig03.ps]{Map of the interstellar polarization in the
vicinity of GW~Ori. The stars selected to calculate the IS polarization
are within 200~pc of this target. \label{Fig-Polis_GWOri}}

\newpage
\scalebox{0.75}{\includegraphics{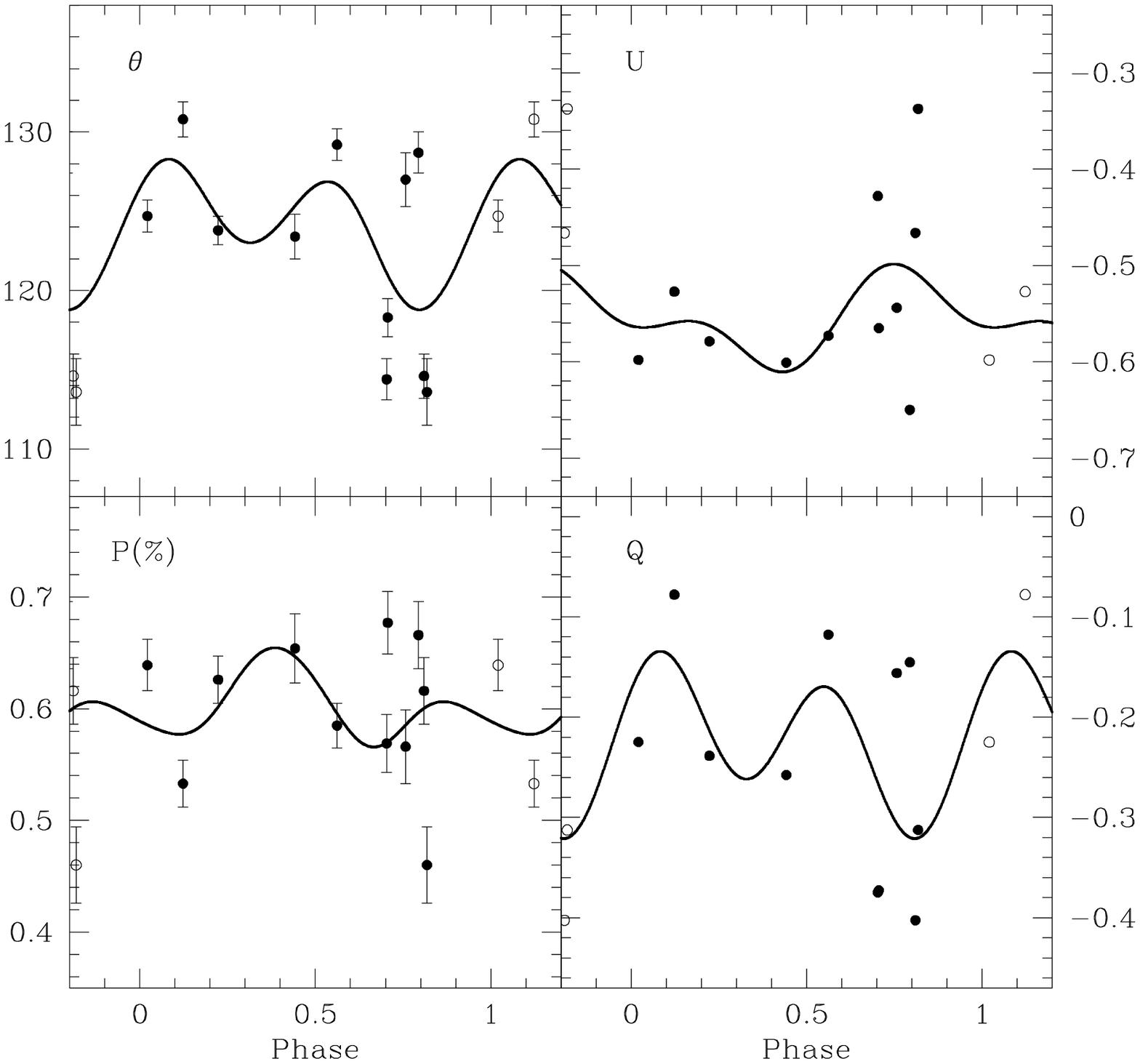}}
\figcaption[Manset4.fig03.ps]{Polarimetric observations of GW~Ori. This
star is polarimetrically variable. \label{Fig-gwori}}

\newpage
\scalebox{0.90}{\includegraphics{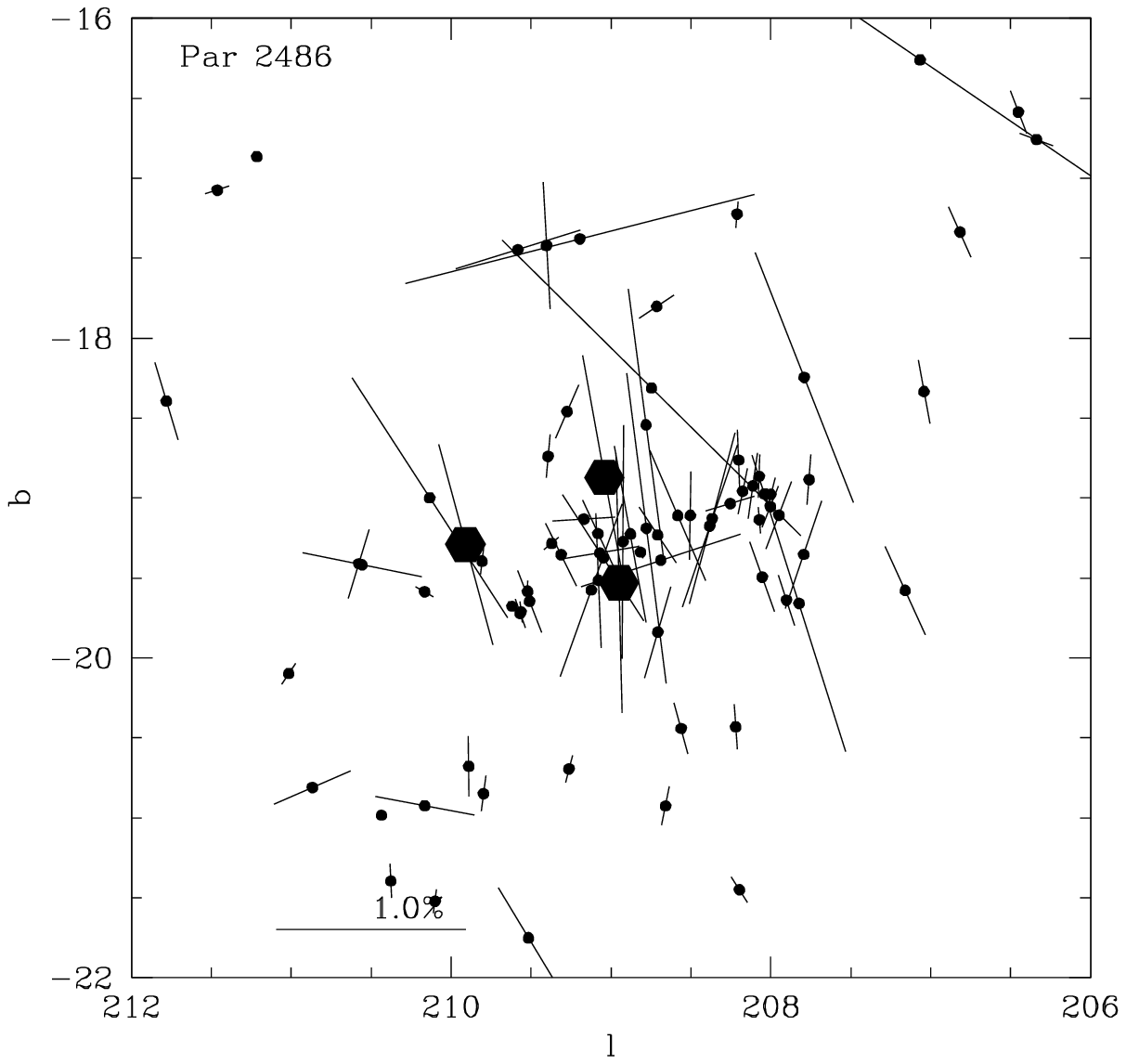}}
\figcaption[Manset4.fig03.ps]{Map of the interstellar polarization in
the vicinity of Par~2486 (at the center of the map), Par~1540 (below
center), and Par~2494 (left). The stars selected to calculate the IS
polarization are within 235~pc of those
targets. \label{Fig-Polis_Parenago}}

\newpage
\scalebox{0.75}{\includegraphics{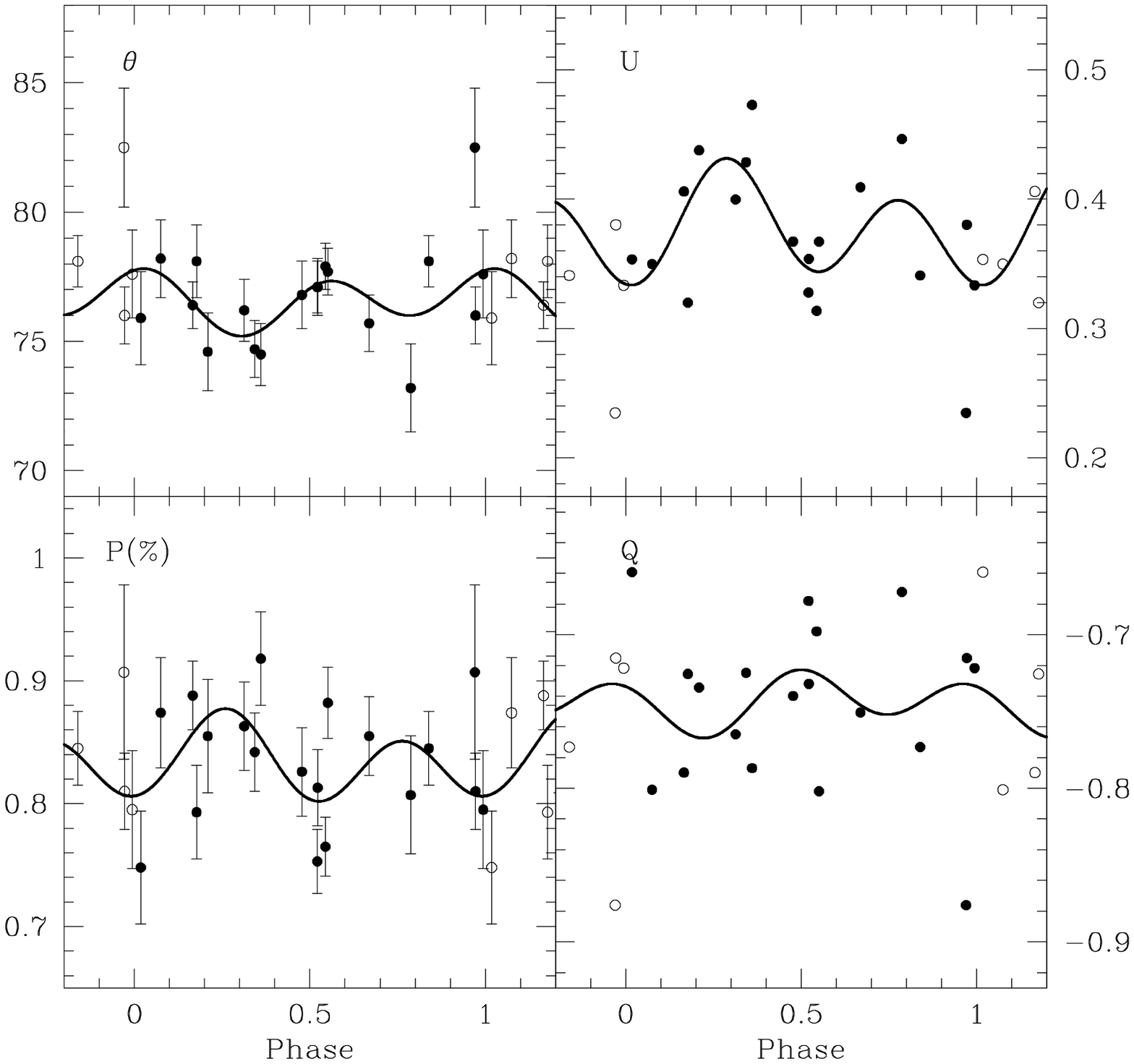}}
\figcaption[Manset4.fig03.ps]{Polarimetric observations of Par~1540; one
point with a rather high polarization of $\approx 1.1$\% was
removed. Clearly, this binary is variable, but there is a lot of
scatter, and periodic variations are not clear. \label{Fig-p1540a}}

\newpage
\scalebox{0.75}{\includegraphics{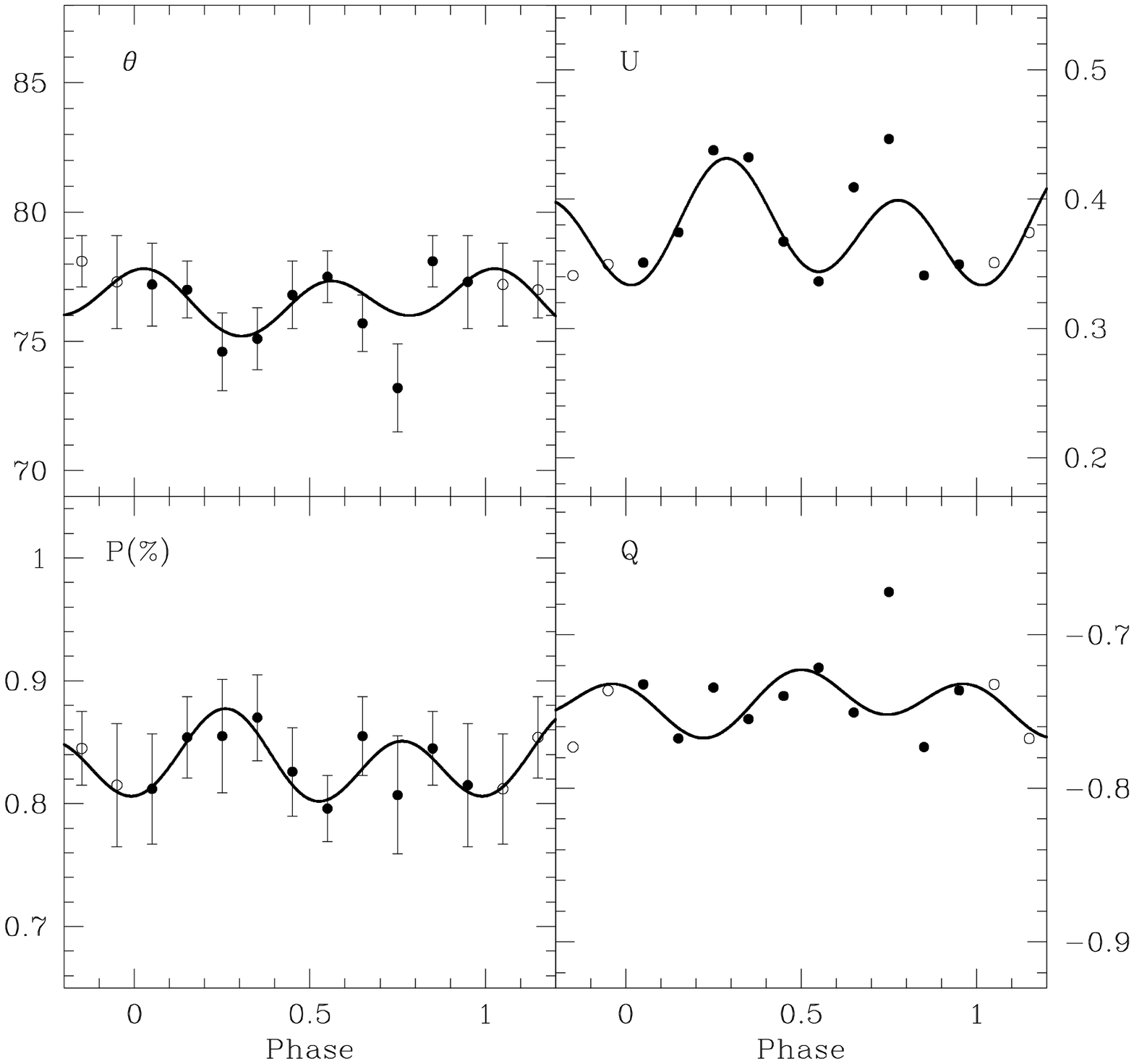}}
\figcaption[Manset4.fig03.ps]{Polarimetric observations of
Par~1540. Data have been binned in phase, to reveal small amplitude
periodic variations. \label{Fig-p1540b}}

\newpage
\scalebox{0.75}{\includegraphics{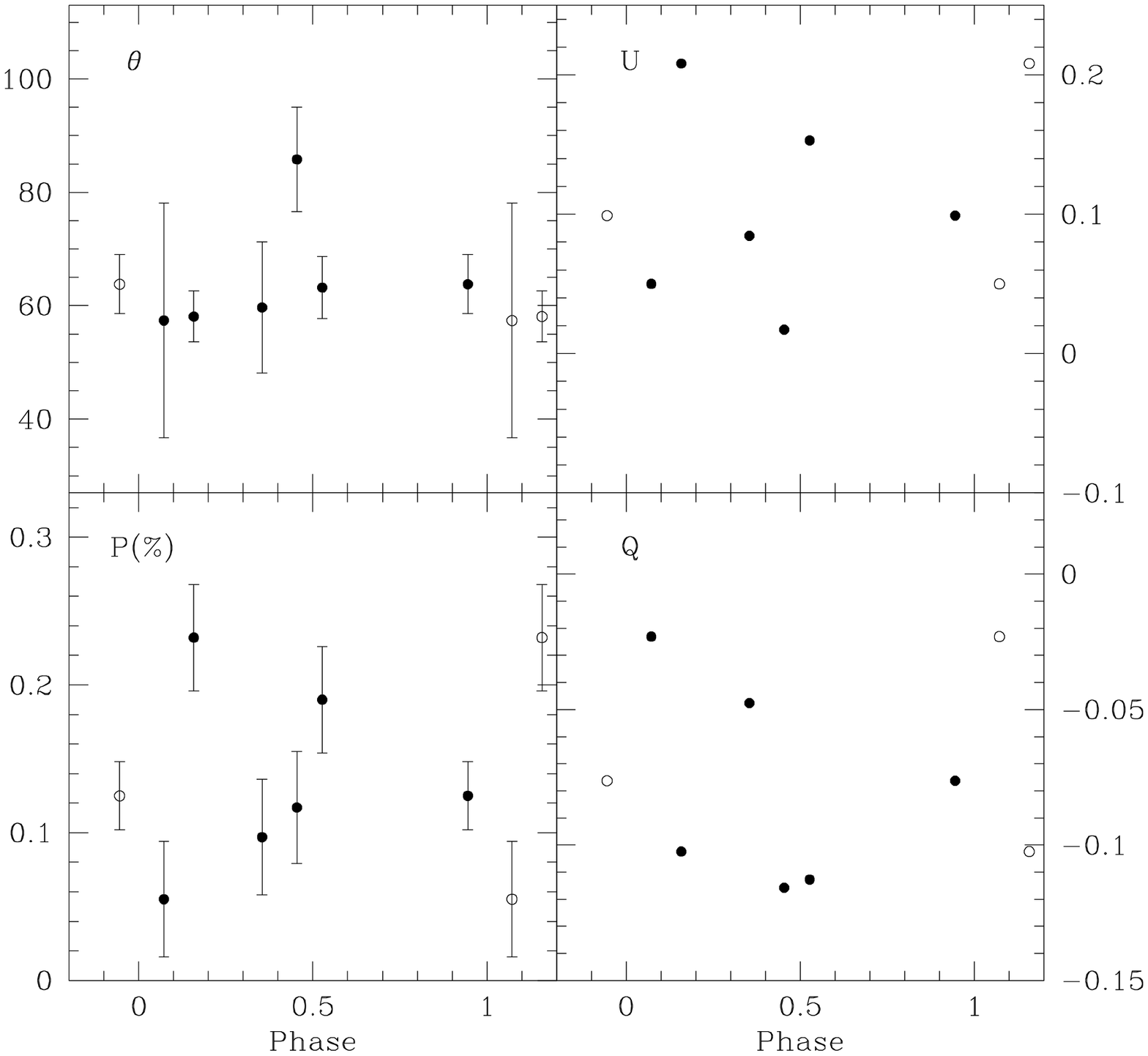}}
\figcaption[Manset4.fig03.ps]{Polarimetric observations of
Par~2486. \label{Fig-p2486}} 

\newpage
\scalebox{0.9}{\includegraphics{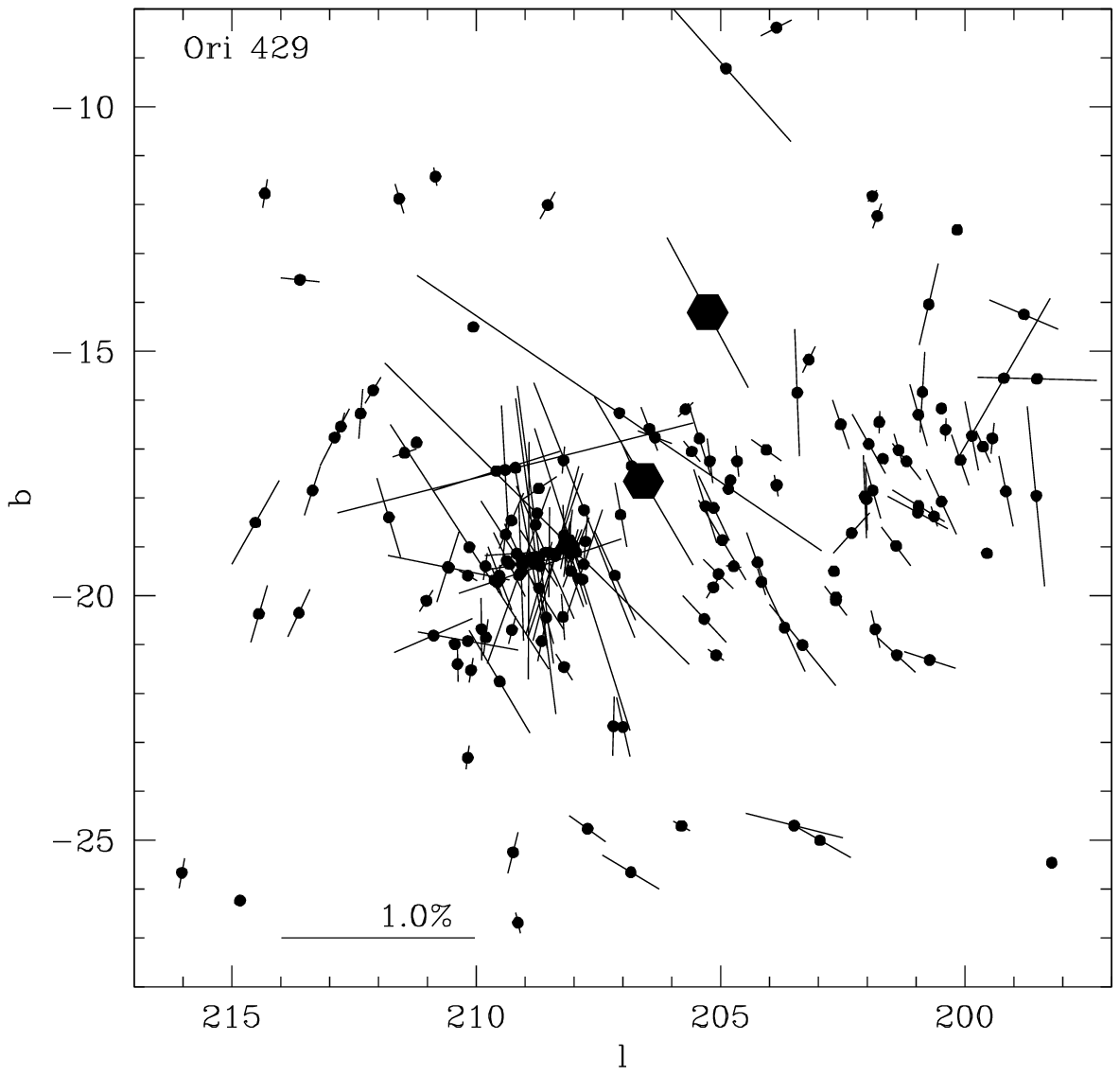}}
\figcaption[Manset4.fig03.ps]{Map of the interstellar polarization in
the vicinity of Ori~429 (at the center of the map) and Ori~569 (above
center). The stars selected to calculate the IS polarization are within
235~pc of those targets. \label{Fig-Polis_ORI429_569}}

\newpage
\scalebox{0.75}{\includegraphics{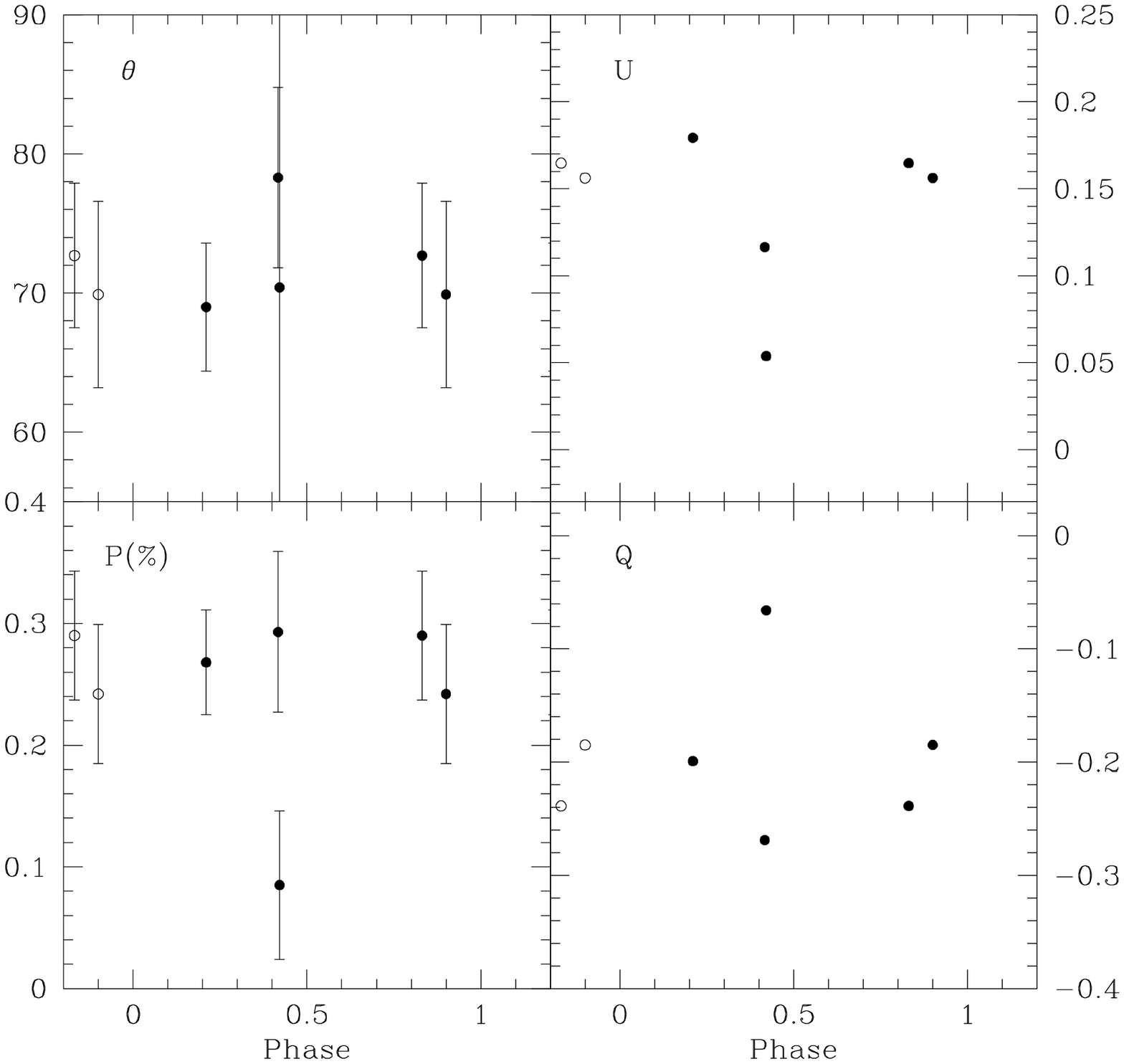}}
\figcaption[Manset4.fig03.ps]{Polarimetric observations of
Ori~429. \label{Fig-ori429}}

\newpage
\scalebox{0.75}{\includegraphics{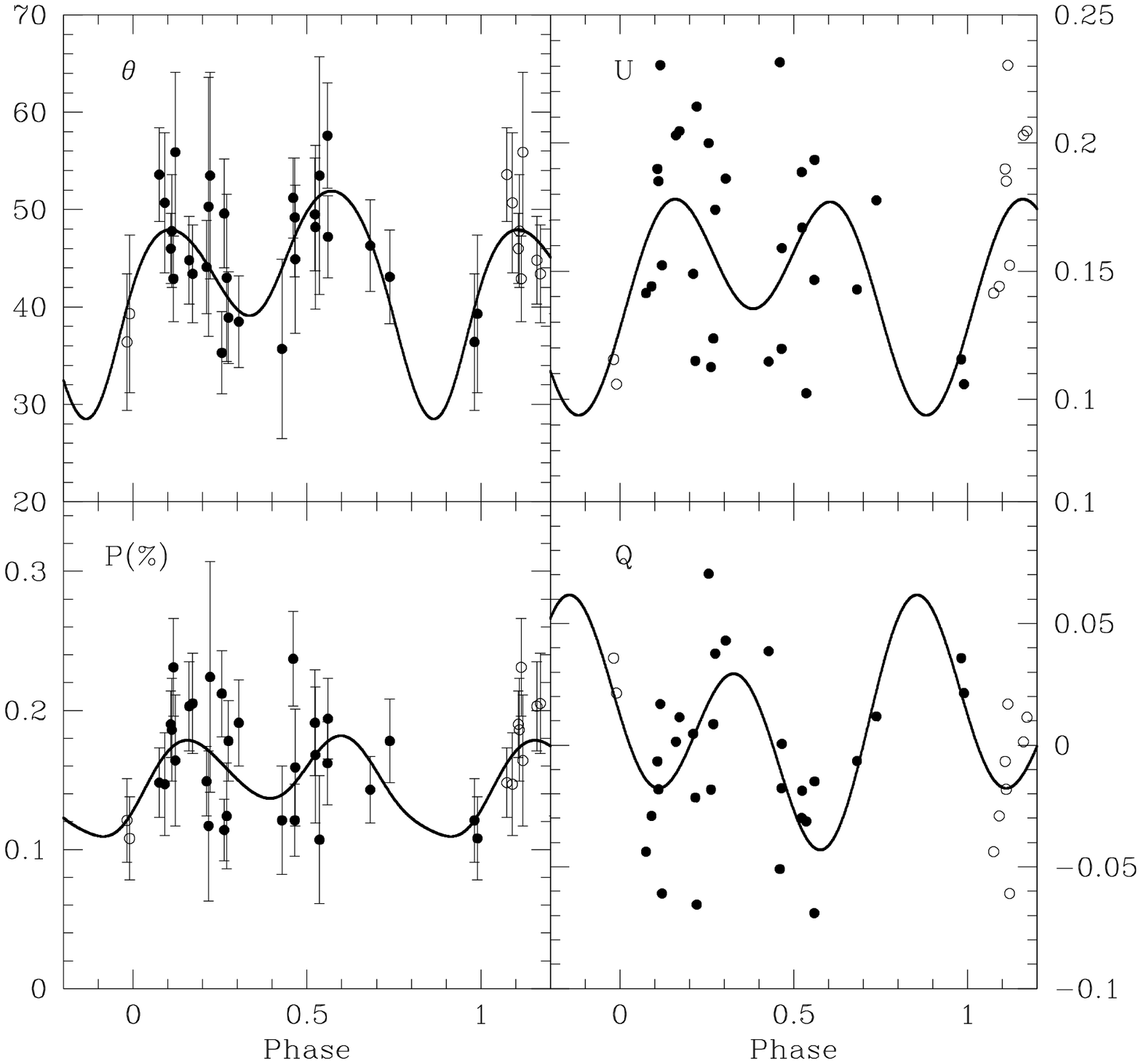}}
\figcaption[Manset4.fig03.ps]{Polarimetric observations of Par~2494,
which was determined to be constant in polarization. The first
observation, taken in 1995 August, is not shown, since its polarization
level and position angle are different from the rest of the
data.\label{Fig-p2494a}} 

\newpage
\scalebox{0.75}{\includegraphics{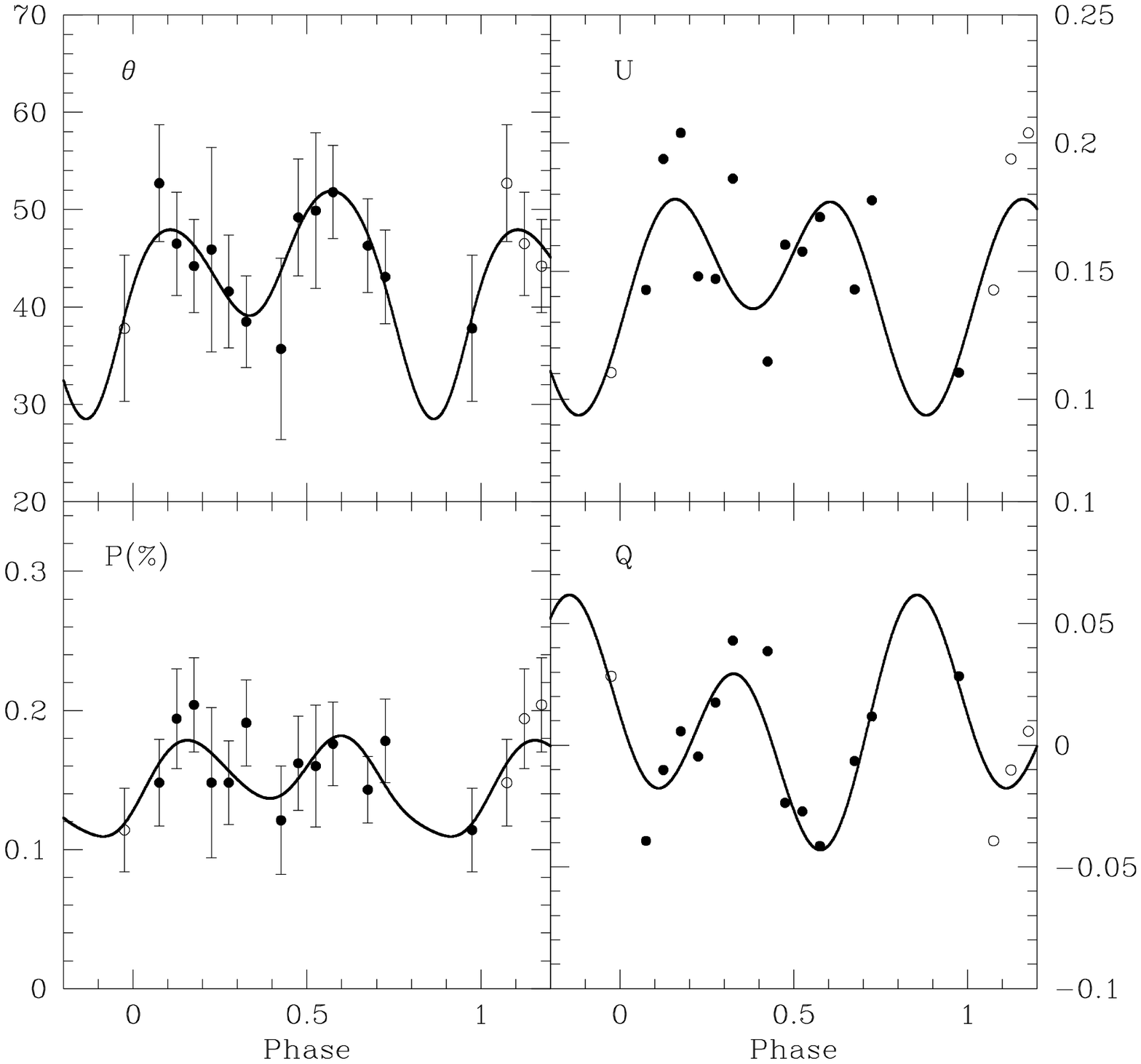}}
\figcaption[Manset4.fig03.ps]{Binned data for polarimetric observations
of Par~2494 reveal phased-locked variations in position
angle. \label{Fig-p2494b}}

\newpage
\scalebox{0.75}{\includegraphics{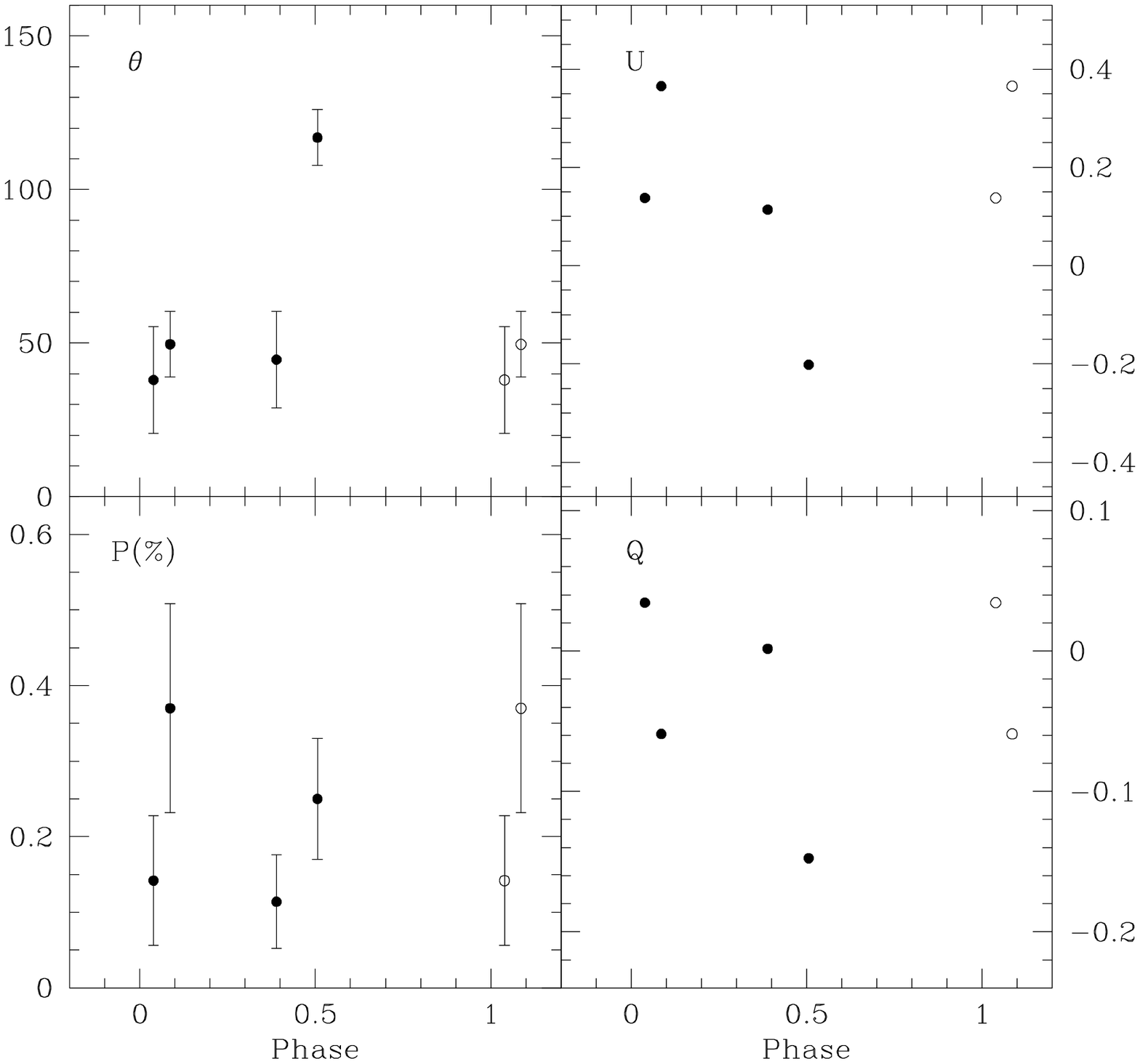}}
\figcaption[Manset4.fig03.ps]{Polarimetric observations of
Ori~569. \label{Fig-ori569}}

\newpage
\scalebox{0.9}{\includegraphics{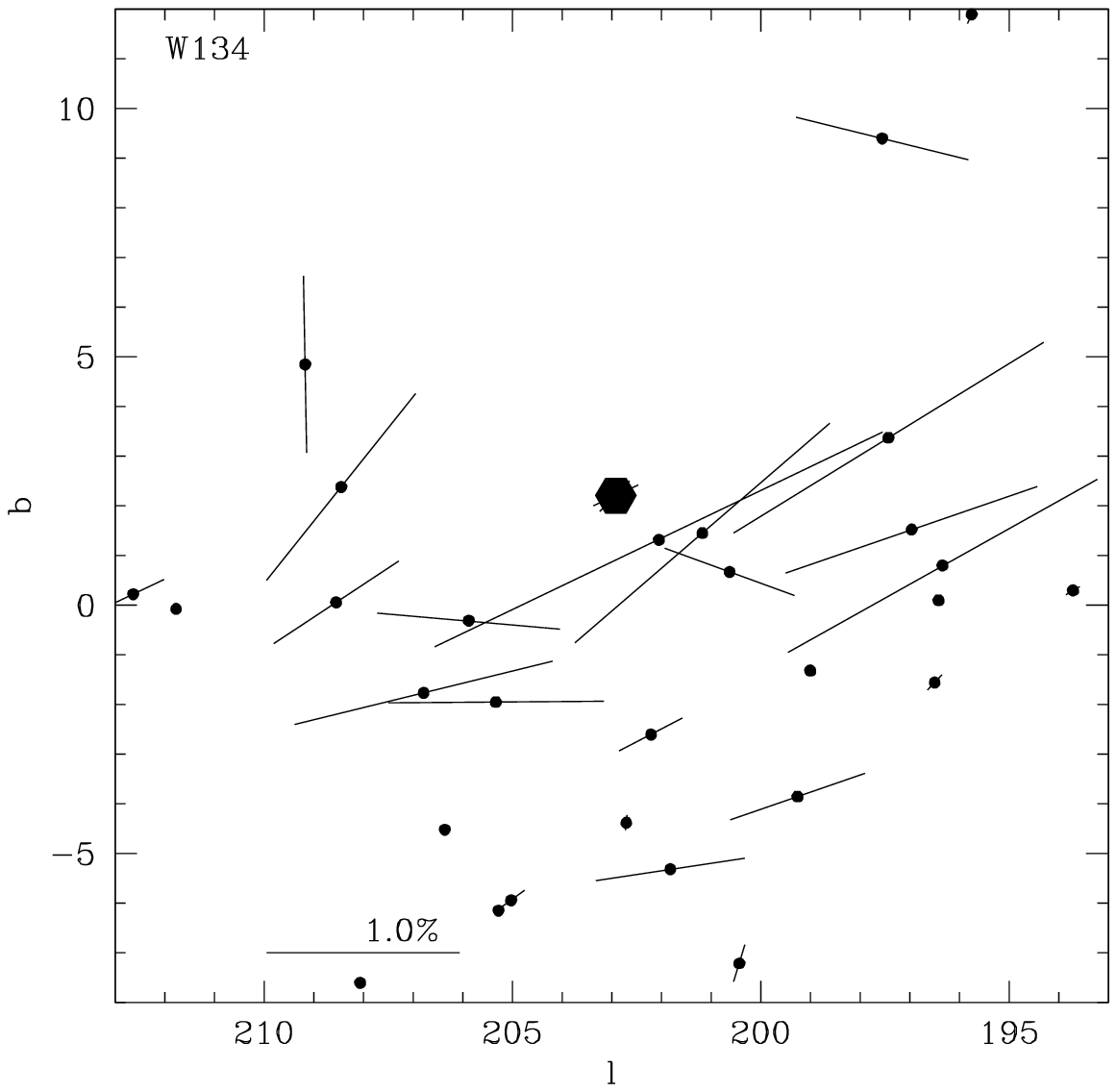}}
\figcaption[Manset4.fig03.ps]{Map of the polarization in the
vicinity of VSB~126 and W~134 (both at the center of the map). The stars
selected to calculate the IS polarization are within 350~pc of those
targets. \label{Fig-Polis_W134_VSB126}}

\newpage
\scalebox{0.75}{\includegraphics{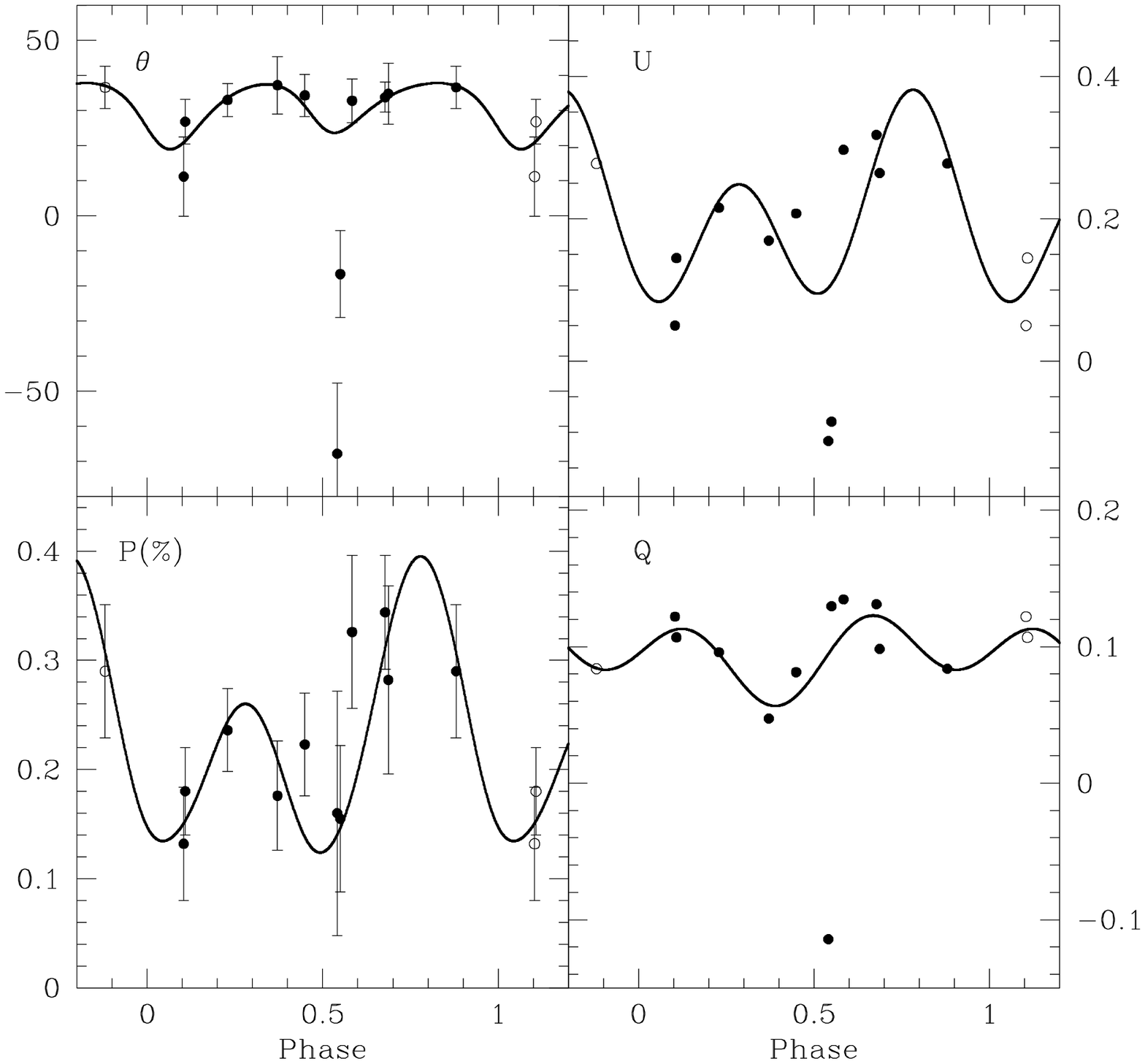}}
\figcaption[Manset4.fig03.ps]{Polarimetric observations of
W~134. \label{Fig-w134}}

\newpage
\scalebox{0.75}{\includegraphics{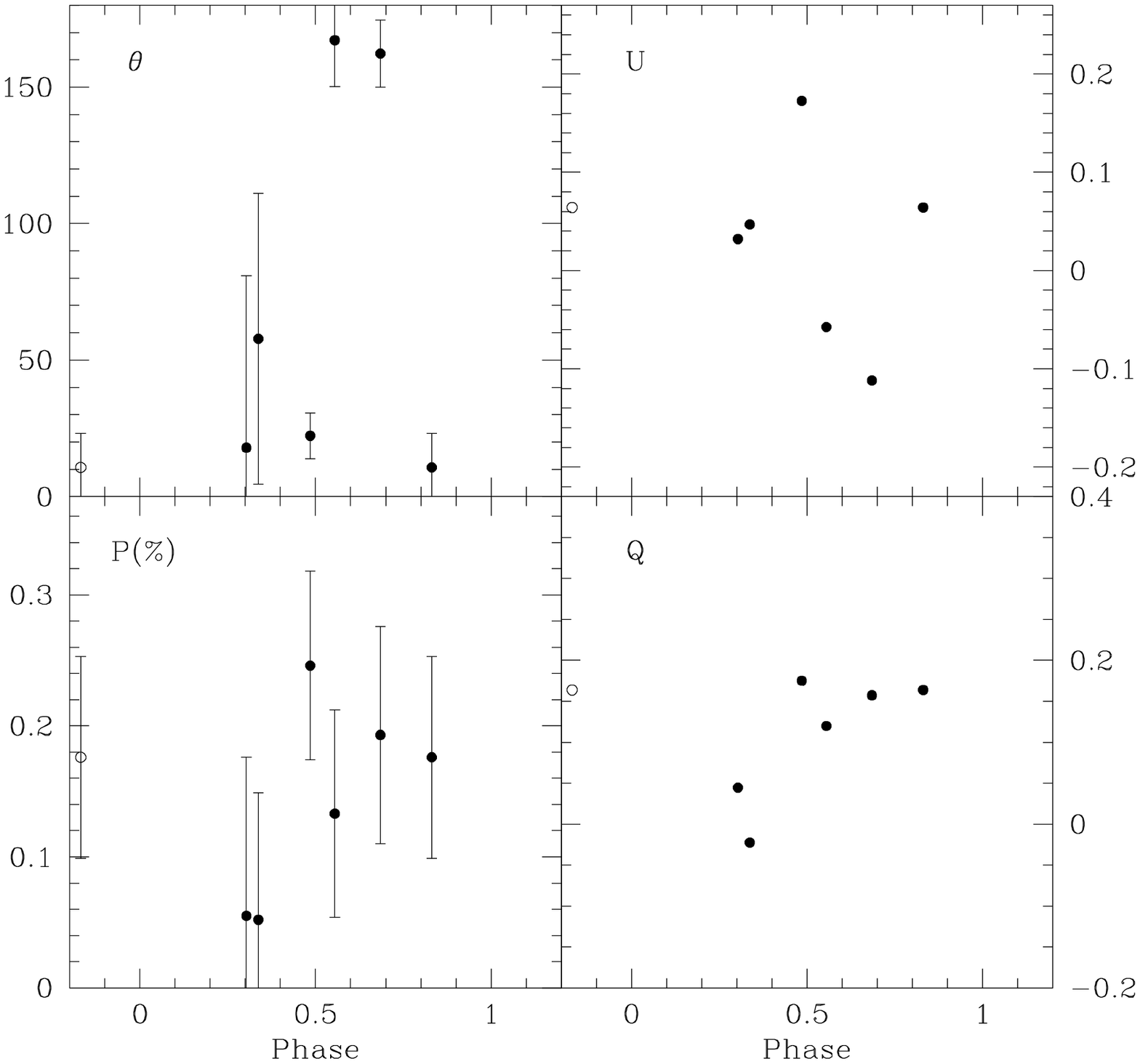}}
\figcaption[Manset4.fig03.ps]{Polarimetric observations of
VSB~126. \label{Fig-vsb126}}


\newpage

\begin{deluxetable}{lrlccl}
\rotate         
\tabletypesize{\scriptsize}
\tablewidth{0pt}        
\tablecaption{Identification, coordinates, and location of the PMS
binaries \label{Tab-Coord}}
\tablehead{
\colhead{Star} & \colhead{HBC\tablenotemark{1}} & \colhead{Other Names} &
\colhead{$\alpha$\tablenotemark{2}} & \colhead{$\delta$\tablenotemark{2}} &
\colhead{Location}\\ 
\colhead{} & \colhead{} & \colhead{} & \colhead{(2000.0)} &
\colhead{(2000.0)} & \colhead{}}
\startdata
V773~Tau &367& HD~283447 & 04 14 13 & $+$28 12 12 & Tau-Aur L1495 \\ 
LkCa~3 &368&\nodata & 04 14 48 & $+$27 52 35 & Tau-Aur L1495\\ 
V826~Tau &400& TAP~43, FK~1 & 04 32 15 & $+$18 01 42 & Tau-Aur L1551\\ 
UZ~Tau~E/W &52 & CoKu UZ Tau f & 04 32 43 & $+$25 52 31 & Tau-Aur B217\\ 
DQ~Tau &72 & IRAS~04439$+$1654 &04 46 52 & $+$16 59 54 & Tau-Aur L1558\\
NTTS~045251+3016&427& TAP~57, V397~Aur &04 56 02 & $+$30 21 03 
         & Tau-Aur L1517\\ 
GW~Ori &85 & HD~244138, BD~$+$11~819 & 05 29 08 & $+$11 52 13 
         & $\lambda$ Ori Assoc. B225\\
Par 1540 &447& NGC~1977~334 &05 34 41 & $-$05 24 36 & Trapezium \\
Par 2486 &\nodata & NGC~1977~1060, BD~$-$05~1340 &05 37 10 & $-$05 10 36 
         & Trapezium \\
Ori~429   &\nodata &\nodata & 05 34 41\tablenotemark{3} & $-$02 33
54\tablenotemark{3}
        & \nodata\\
Par~2494 &487& NGC~1977~1069, BD~$-$06~1258 &05 37 09 & $-$06 06 12 &
        Trapezium \\ 
Ori~569 &\nodata &\nodata & 05 44 29\tablenotemark{3} & $-$00 10
30\tablenotemark{3}  & L1630 (M78)\\
W~134 &536& NGC~2264~134, VAS~92, VSB~92 & 06 40 59 & $+$09 55 20 
         & NGC 2264\\
VSB~126 &\nodata & NGC~2264~169, NGC~2264 VAS~126 & 06 41 08 & $+$09 44 03 
         & NGC 2264
\tablenotetext{1}{HBC numbers come from the Herbig and Bell Catalog
(Herbig \& Bell 1988).}
\tablenotetext{2}{All coordinates except for Ori~429 and Ori~569 come
from SIMBAD; coordinates for Ori~429 and Ori~569: F. Walter 1995, private
communication.}
\tablenotetext{3}{Equinox 1950.0 (F. Walter 1995, private communication).}
\enddata
\end{deluxetable}

\pagebreak \clearpage


\begin{deluxetable}{llcccccrlcccrc}
\rotate
\tabletypesize{\scriptsize}
\tablewidth{0pt}
\tablecaption{Spectroscopic and orbital information for the PMS binaries
\label{Tab-Spectro}} 
\tablehead{
\colhead{Star\tablenotemark{1}} & \colhead{Spectral} & \colhead{Ref.} & 
\colhead{Type\tablenotemark{2}} & \colhead{Ref.} & 
\colhead{sgl/\tablenotemark{3}} 
& \colhead{Ref.} & \colhead{Period} & \colhead{Ecc.} & \colhead{Ref.} & 
\colhead{Inc.} & \colhead{Ref.} & \colhead{Dist.} & \colhead{Ref.} \\
\colhead{}     & \colhead{Type} & \colhead{} & 
\colhead{} & \colhead{} &
\colhead{dbl} & \colhead{} &
\colhead{(d)} & \colhead{} & \colhead{} & 
\colhead{($\arcdeg$)} &\colhead{} & 
\colhead{(pc)} & \colhead{}}
\startdata
V773~Tau (3?) & K3 V(Li) &1& WT &1& dbl&2& 51.075 & 0.267 &2
        & $\gtrsim 66$ &2 &170&2\\
LkCa~3 (3) & M1 V(Li) &1&\nodata &\nodata & sgl&2& 12.941 & 0.20
&3&\nodata &\nodata &140&2\\ 
V826~Tau (2) & K5-K7 + &4& WT &1& dbl&2& 3.88776 & 0.0 &3& 7-13 
        & 5, 6 &160&2\\
             & K5-K7 &&&&&&&&&&&&\\
UZ~Tau~E (2) & M1,3:V(Li) &1& CT &1& sgl& 7&19.1 & 0.28 & 7&\nodata
&\nodata &140&\nodata\\ 
DQ~Tau  & M0,1 V(Li) &1& CT &1& dbl& 8&15.8 & 0.58 & 8&\nodata &\nodata
&\nodata&\nodata\\ 
NTTS~045251+3016 & K7(Li) &1& WT &1& sgl& 9&2530 & 0.48 &3&\nodata
&\nodata&\nodata&\nodata\\ 
GW~Ori (3) & G5 (Li) &1& CT &1& sgl&10&241.9 & 0.04 &3&15-90 &\nodata&400&10 \\
Par~1540 (2) & K4-K5 &4& WT &4& dbl&11&33.73 & 0.12 &3&\nodata
&\nodata&470&12\\ 
Par~2486 (2) & \nodata &\nodata &\nodata &\nodata & dbl&3&5.1882 & 0.161
&3&\nodata &\nodata&470&\nodata\\ 
Ori~429 (2)  & K3 + &4& WT &4& dbl&4&7.46 & 0.27 &3&\nodata &\nodata&470&4\\
             & K3-K5 &&&&&&&&&&&&\\
Par~2494 (2) & K0 IV(Li) &1&\nodata &\nodata & dbl&\nodata&19.4815 &
0.262 &3&\nodata &\nodata&470&\nodata\\ 
Ori~569 (3) & K4-K5 + &4& WT &4& dbl&4&4.25 & 0.0 &3&\nodata &\nodata&470&4\\
            & K4-K5 &&&&&&&&&&&&\\
W~134 (2) & G5 V &1&\nodata &\nodata & dbl&\nodata&6.3532 & $<$0.01 &13&
46, 63&13& 700&\nodata\\ 
VSB~126 (2) &\nodata  &\nodata &\nodata &\nodata & sgl&3&12.924 & 0.18
&3&\nodata &\nodata&700&\nodata
\tablenotetext{1}{The numbers in parenthesis after each object indicate
the number of stars in each system.}
\tablenotetext{2}{Type of PMS star: CT (classical TTS), WT
(weak-line TTS)}
\tablenotetext{3}{Single-line (sgl) or double-line (dbl) spectroscopic binary.}
\tablerefs{(1) Herbig \& Bell 1988 (HBC catalog) and references cited;
(2) Welty 1995;
(3) Mathieu 1994 and references cited;
(4) Lee, Mart\'{\i}n, \& Mathieu 1994;
(5) Reipurth et al. 1990;
(6) Mundt et al. 1983;
(8) Stassun et al. 1996;
(7) Mathieu, Mart\'{\i}n, \& Maguzzu 1996;
(9) Walter et al. 1988;
(10) Mathieu, Adams, \& Latham 1991;
(11) Marshall \& Mathieu 1988;
(12) Jones \& Walker 1988;
(13) Padgett \& Stapelfeldt 1994}
\enddata
\end{deluxetable}

\pagebreak \clearpage

\begin{deluxetable}{lccccccccccccc}
\rotate
\tabletypesize{\scriptsize}
\tablewidth{0pt}
\tablecaption{Average observed polarization, origin of the polarization,
and estimate of the intrinsic polarization for the PMS binaries
\label{Tab-averpol}}  
\tablehead{
\colhead{Star} & \colhead{$P_{\rm ave}$\tablenotemark{1}} & 
\colhead{$\theta_{\rm ave}$\tablenotemark{1}} & 
\colhead{$N_{\rm obs}$} & \colhead{Origin of} &
\colhead{$P_{\rm IS}$\tablenotemark{3}} & 
\colhead{$\sigma(P_{\rm IS})$} &
\colhead{$\theta_{\rm IS}$\tablenotemark{3}} & 
\colhead{$\sigma(\theta_{\rm IS})$ }&
\colhead{$N_{\rm IS}$} & \colhead{Radius} & \colhead{Distance} &
\colhead{$P_{\star}$\tablenotemark{4}} & 
\colhead{$\theta_{\star}$\tablenotemark{4}} \\
\colhead{} & \colhead{(\%)} & \colhead{(\arcdeg)} & 
\colhead{} & \colhead{Polarization\tablenotemark{2}} &
\colhead{(\%)} & \colhead{(\%)} & 
\colhead{(\arcdeg)} & \colhead{(\arcdeg)} &
\colhead{} & \colhead{(\arcdeg)} & \colhead{($\pm$ pc)} &
\colhead{(\%)} & \colhead{(\arcdeg)}}
\startdata
V773~Tau & 0.35 & 88 & 6  & $\star$ $+$ IS? &
	0.07&0.08&72 &36&24&15&85&0.3 &92\\
LkCa~3 & 0.05 & 76 & 12 & IS $+$ $\star$? &
	0.11&0.07&8  &18&23&15&70&0.15&91  \\
V826~Tau & 0.85 & 67 & 11 & $\star$ &
	0.04&0.05&131&32&32&15&80&0.9 &66 \\
UZ~Tau~E/W & 0.80 & 16 & 2  & $\star$ $+$ IS  &
	0.11&0.05&17 &14&26&15&70&0.7 & 15\\
DQ~Tau & 0.57 & 79 & 1  & $\star$ $+$ IS?&
	0.11&0.07&85 &17&19&12&80&0.5 & 78\\
NTTS~045251+3016& 0.10 & 107 & 3 & $\star$ $+$ IS &
	0.20&0.07&58 & 9&18&15&80&0.2 &136 \\
GW~Ori& 0.61 & 126 & 11& $\star$ $+$ IS? &
	0.04&0.04&41 &30&30&10&200&0.65&126 \\
Par 1540 & 0.83 & 77 & 19 & $\star$ $+$ IS  &
	0.30&0.05&64 &5&28&1.0&235&0.6 & 84\\
Par 2486& 0.14 & 63 & 6  & IS $+$ $\star$ &
	0.38&0.06&73 &5&32&1.0&235&0.25&169 \\
Ori~429 & 0.24 & 72 & 5  & IS&
	0.35&0.07&92 &6&21&2&235&0.2 & 23\\
Par~2494 & 0.16 & 46 & 29 & $\star$ $+$ IS  &
	0.26&0.05&78 & 5&48&2&235&0.2 & 7 \\
Ori~569 & 0.18 & 76 & 4  & $\star$ $+$ IS  &
	0.21&0.04& 90&5&68&6&235&0.1 & 29\\
W~134 & 0.22 & 32 & 11 & $\star$ $+$ IS &
	0.87&0.12&177&4&20&8&350&0.8 & 80\\
VSB~126 & 0.16 & 66 & 6  & $\star$ $+$ IS  &
	0.86&0.12&177&4&20&8&350&1.0 & 84
\tablenotetext{1}{Weighted averages of the observed $P$ and
$\theta$. The averages do not include atypical observations.}
\tablenotetext{2}{Probable origin of the polarization. A $\star$ symbol
indicates intrinsic polarization; IS stands for interstellar
polarization. If IS comes before a $\star$ symbol, the IS component of
the polarization is probably stronger than the intrinsic one.} 
\tablenotetext{3}{Estimate of the IS polarization at the
location of the PMS binary, based on a weighted average of the
polarization of stars in the neighborhood and at similar distance to the
targets. Data taken from the Heiles (2000) catalog.}
\tablenotetext{4}{Intrinsic polarization obtained after
subtracting the estimated IS polarization.}
\enddata
\end{deluxetable}


\pagebreak \clearpage


\begin{deluxetable}{lll}
\tablewidth{0pt}
\tablecaption{Use of the variability tests \label{Tab-UseVarTests}}
\tablehead{
\colhead{Variability Result} & \colhead{Variance and $Z$ Tests} & 
\colhead{$\chi ^2$ Test}}
\startdata
Variable & $\ge 3$ PR & $\ge 3$ PR \\
         & $\ge 3$ PR & 2 PR, for same Stokes parameter \\
         &            & as Variance and $Z$ tests\\
        & $Z_Q$ and $Z_U$ PR & $\ge 3$ PR \\
        & $\sigma_Q$ and $Z_Q$ PR & $\ge 3$ PR \\
        & $\sigma_U$ and $Z_U$ PR & $\ge 3$ PR \\ \\
Suspected Var. & $\ge 3$ PR & 1 PR\\ \\
Possibly Const. & $\ge 3$ PR & 0 PR\\ \\
Constant & 1 or 2 PR for the $\sigma$ test, & 0 PR\\
         & none for the $Z$ test & 
\tablecomments{This table indicates the number of positive results (PR)
from the variability tests needed to reach the different variability
classification.} 
\enddata
\end{deluxetable}

\pagebreak \clearpage

\begin{deluxetable}{lccccc}
\tablewidth{0pt}
\tablecaption{Amplitude of the polarimetric variations\tablenotemark{1}
\label{Tab-AmpVar}}
\tablehead{
\colhead{Star} & \colhead{$\Delta P$} & \colhead{$\Delta \theta$} &
\colhead{$\Delta Q$} & \colhead{$\Delta U$} & \colhead{$N_{\rm obs}$}\\
\colhead{} & \colhead{(\%)} & \colhead{$(\arcdeg) $} & \colhead{(\%)} &
\colhead{(\%)} & \colhead{}} 
\startdata
V773~Tau        & 0.28 & 14.8 & 0.26 & 0.20 & 6\\
LkCa~3          & 0.13 & 97.9 & 0.13 & 0.08 & 12\\
V826~Tau        & 0.17 & 11.6 & 0.23 & 0.27 & 11\\
UZ~Tau~E/W      & 0.49 & 46.1 & 1.15 & 0.54 & 2\\
NTTS~045251+3016& 0.10 & 28.5 & 0.07 & 0.11 & 3\\
GW~Ori          & 0.22 & 17.2 & 0.33 & 0.31 & 11\\
Par 1540        & 0.17 & 9.3 & 0.22 & 0.24 & 19\\
Par 2486        & 0.18 & 28.4 & 0.09 & 0.19 & 6\\
Ori~429         & 0.21 & 9.3 & 0.20 & 0.13 & 5\\
Par~2494        & 0.13 & 22.3 & 0.14 & 0.13 & 29\\
Ori~569         & 0.26 & 78.9 & 0.18 & 0.57 & 4\\
W~134           & 0.21 & 128.8 & 0.25 & 0.43 & 11\\
VSB~126         & 0.19 & 156.6 & 0.20 & 0.29 & 6
\tablenotetext{1}{Difference between the minimum and maximum values of
the polarization, excluding, in some cases, very atypical observations.}
\enddata
\end{deluxetable}

\pagebreak \clearpage

\begin{deluxetable}{lccccccc}
\tablewidth{0pt}
\tablecaption{Results of the variability tests \label{Tab-VarDetails}}
\tablehead{
\colhead{Star} & \colhead{$N_{\rm obs}$} & \colhead{} & 
\colhead{$\sigma_{\rm sample}$} & \colhead{$\sigma_{\rm mean}$} &
\colhead{$Z \pm \sigma{Z}$} & \colhead{$P{\chi^2}$} &
\colhead{$P{\chi^2}$}\\
\colhead{} & \colhead{} & \colhead{} & 
\colhead{} & \colhead{} &
\colhead{} & \colhead{$1\sigma$} &
\colhead{$1.5\sigma$}}
\startdata
V773~Tau  &6&$Q$& 0.0912 & 0.0122 & 3.36 0.32 & 1.00 & 1.00\\
          & &$U$& 0.0777 & 0.0122 & 2.60 0.32 & 1.00 & 0.97\\
LkCa~3   &12&$Q$& 0.0314 & 0.0090 & 0.95 0.21 & 0.45 & 0.04\\
         &  &$U$& 0.0277 & 0.0090 & 0.84 0.21 & 0.27 & 0.02\\
V826~Tau &11&$Q$& 0.0659 & 0.0104 & 1.69 0.22 & 0.99 & 0.76\\
         &  &$U$& 0.0753 & 0.0104 & 2.05 0.22 & 1.00 & 0.95\\
UZ~Tau~E/W&2&$Q$& 0.8147 & 0.0274 & 21.03 0.71 & 1.00 & 1.00\\
          & &$U$& 0.3819 & 0.0274 & 9.86 0.71 & 1.00 & 1.00\\
NTTS~045251+3016 &3&$Q$& 0.0399 & 0.0281 & 0.68 0.50 & 0.29 & 0.14\\
                 & &$U$& 0.0573 & 0.0281 & 1.10 0.50 & 0.70 & 0.42\\
GW~Ori   &11&$Q$     & 0.1119 & 0.0077 & 4.47 0.22 & 1.00 & 1.00 \\
         &  &$U$     & 0.0901 & 0.0077 & 3.04 0.22 & 1.00 & 1.00\\
Par 1540  &19&$Q$    & 0.0529 & 0.0078 & 1.34 0.17 & 0.98 & 0.28\\
          &  &$U$    & 0.0564 & 0.0078 & 1.39 0.17 & 0.99 & 0.38\\
Par 2486  & 6&$Q$    & 0.0378 & 0.0136 & 0.99 0.32 & 0.58 & 0.18\\
          &  &$U$    & 0.0694 & 0.0136 & 1.88 0.32 & 0.99 & 0.84\\
Ori~429 & 5&$Q$      & 0.0776 & 0.0242 & 1.27 0.35 & 0.83 & 0.42\\
        &  &$U$      & 0.0506 & 0.0242 & 0.92 0.35 & 0.48 & 0.15\\
Par~2494 &29&$Q$     & 0.0342 & 0.0059 & 1.01 0.13 & 0.56 & 0.01 \\
         &  &$U$     & 0.0386 & 0.0059 & 1.17 0.13 & 0.90 & 0.05\\
Ori~569 & 4&$Q$      & 0.0800 & 0.0407 & 1.01 0.41 & 0.61 & 0.28\\
        &  &$U$      & 0.2331 & 0.0407 & 2.47 0.41 & 1.00  & 0.95\\
W~134  &11&$Q$       & 0.0706 & 0.0162 & 0.81 0.22 & 0.18 & 0.01\\
       &  &$U$       & 0.1483 & 0.0162 & 2.13 0.22 & 1.00 & 0.97\\
VSB~126 & 6&$Q$      & 0.0790 & 0.0345 & 0.88 0.32 & 0.40 & 0.10\\
        &  &$U$      & 0.0994 & 0.0345 & 1.29 0.32 & 0.86 & 0.41\\
\enddata
\end{deluxetable}

\pagebreak \clearpage

\begin{deluxetable}{ll}
\tablewidth{0pt}
\tablecaption{Classification of the observed PMS binaries,
according to their variability\label{Tab-Var}} 
\tablehead{
\colhead{Variability Classification} & {PMS Binary}}
\startdata
Variable & V773~Tau (6), V826~Tau (11), GW~Ori (11)\\
         & Par~1540 \tablenotemark{1} (19), W~134 (11)\\  \\
Suspected variable & Par~2486 (6), Ori~569 (4)\\ \\
Constant & LkCa~3 (12), NTTS~045251+3016 (3), Ori~429 (5),\\
        & Par~2494 \tablenotemark{1} (29), VSB~126 (6)
\tablenotetext{1}{These stars were sometimes observed to have very
different polarization and/or position angle values (well above or below
the majority of the data points) so their variability is based on data
excluding those very atypical values.}
\tablecomments{The number of observations used for the variability tests
is indicated in parentheses.}
\enddata
\end{deluxetable}

\pagebreak \clearpage

\begin{deluxetable}{lcccccc}
\tablewidth{0pt}
\tablecaption{Polarization data for V773 Tau \label{Tab-v773tau}}
\tablehead{
\colhead{UT Date} & \colhead{JD} & \colhead{Phase\tablenotemark{1}} & 
\colhead{P} & \colhead{$\sigma(P)$} &
\colhead{$\theta$} & \colhead{$\sigma(\theta)$}\\
\colhead{} & 2400000.0$+$ & \colhead{} &
\colhead{(\%)} & \colhead{(\%)} &
\colhead{(\arcdeg)} & \colhead{(\arcdeg)}}
\startdata
1998 Jan 22 & 50835.493 & 0.316 & 0.189  & 0.025   &  94.9  &   3.8\\     
1998 Dec 14 & 51161.509 & 0.699 & 0.469  & 0.058   &  80.1  &   3.6\\
1999 Feb 6  & 51215.615 & 0.179 & 0.397  & 0.033   &  85.6  &   2.4\\        
1999 Feb 9  & 51218.668 & 0.239 & 0.369  & 0.023   &  93.0  &   1.7\\        
1999 Feb 15 & 51224.632 & 0.355 & 0.385  & 0.033   &  86.1  &   2.5\\        
1999 Mar 10 & 51247.586 & 0.384 & 0.432  & 0.031   &  82.6  &   2.0
\tablenotetext{1}{Calculated with the ephemeris $2449900.0+51.075E$
(period from Jensen \& Mathieu 1997).}
\enddata
\end{deluxetable}

\pagebreak \clearpage

\begin{deluxetable}{lcccccc}
\tablewidth{0pt}
\tablecaption{Polarization data for LkCa 3 \label{Tab-lkca3}}
\tablehead{
\colhead{UT Date} & \colhead{JD} & \colhead{Phase\tablenotemark{1}} & 
\colhead{P} & \colhead{$\sigma(P)$} &
\colhead{$\theta$} & \colhead{$\sigma(\theta)$}\\
\colhead{} & 2400000.0$+$ & \colhead{} &
\colhead{(\%)} & \colhead{(\%)} &
\colhead{(\arcdeg)} & \colhead{(\arcdeg)}}
\startdata
1996 Jan 3 & 50085.630 &  0.344  & 0.031  & 0.043  &  119.4  &  38.9\\
1996 Jan 5 & 50087.506 &  0.489  & 0.065  & 0.034  &   68.3  &  15.0\\
1996 Jan 7 & 50089.576 &  0.649  & 0.050  & 0.029  &   39.8  &  16.5\\
1996 Aug 24 & 50319.767 &  0.436  & 0.000  & 0.038  &  137.7  &  57.3\\
1996 Aug 25 & 50320.779 &  0.515  & 0.043  & 0.025  &   91.6  &  16.7\\
1996 Sep 4 & 50330.745 &  0.285  & 0.129  & 0.034  &   78.1  &   7.6\\
1996 Sep 7 & 50333.708 &  0.514  & 0.050  & 0.031  &   69.6  &  17.8\\
1996 Oct 12 & 50368.676 &  0.216  & 0.040  & 0.024  &   69.2  &  17.0\\
1997 Feb 9 & 50488.511 &  0.476  & 0.045  & 0.028  &  101.3  &  18.1\\
1997 Feb 14 & 50493.594 &  0.869  & 0.036  & 0.031  &   98.1  &  24.8\\
1997 Oct 18 & 50739.896 &  0.901  & 0.059  & 0.032  &   73.9  &  15.4\\
1997 Oct 26 & 50747.784 &  0.511  & 0.054  & 0.048  &   67.6  &  25.6
\tablenotetext{1}{Calculated with the ephemeris $2449900.0+
12.941E$ (period from Mathieu 1994).}
\enddata
\end{deluxetable}

\pagebreak \clearpage

\begin{deluxetable}{lcccccc}
\tablewidth{0pt}
\tablecaption{Polarization data for V826 Tau \label{Tab-v826tau}}
\tablehead{
\colhead{UT Date} & \colhead{JD} & \colhead{Phase\tablenotemark{1}} & 
\colhead{P} & \colhead{$\sigma(P)$} &
\colhead{$\theta$} & \colhead{$\sigma(\theta)$}\\
\colhead{} & 2400000.0$+$ & \colhead{} &
\colhead{(\%)} & \colhead{(\%)} &
\colhead{(\arcdeg)} & \colhead{(\arcdeg)}}
\startdata
1994 Sep 22& 49709.744 & 0.147 & 0.739 & 0.039 &   73.2 &   1.5\\
1994 Dec 23& 50084.583 & 0.562 & 0.772 & 0.041 &   61.6 &   1.5\\
1996 Jan 2 & 50089.529 & 0.835 & 0.911 & 0.032 &   66.8 &   1.0\\
1996 Aug 25& 50320.846 & 0.334 & 0.878 & 0.027 &   64.8 &   1.0\\
1996 Sep 7 & 50333.730 & 0.648 & 0.899 & 0.042 &   67.3 &   1.3\\
1996 Oct 12& 50368.696 & 0.641 & 0.824 & 0.028 &   66.5 &   1.0\\
1997 Feb 9 & 50488.679 & 0.503 & 0.871 & 0.035 &   68.3 &   1.2\\
1997 Feb 14& 50493.557 & 0.758 & 0.849 & 0.036 &   66.3 &   1.2\\
1997 Sep 9 & 50700.770 & 0.057 & 0.877 & 0.032 &   66.9 &   1.0\\
1997 Oct 26& 50747.754 & 0.142 & 0.748 & 0.050 &   67.3 &   1.9\\
1998 Jan 22& 50835.468 & 0.703 & 0.867 & 0.035 &   69.4 &   1.2
\tablenotetext{1}{Calculated with the ephemeris $2446840.004
+3.887758E$ (Reipurth et al. 1990).}
\enddata
\end{deluxetable}

\pagebreak \clearpage

\begin{deluxetable}{lcccccc}
\tablewidth{0pt}
\tablecaption{Polarization data for GW Ori \label{Tab-gwori}}
\tablehead{
\colhead{UT Date} & \colhead{JD} & \colhead{Phase\tablenotemark{1}} & 
\colhead{P} & \colhead{$\sigma(P)$} &
\colhead{$\theta$} & \colhead{$\sigma(\theta)$}\\
\colhead{} & 2400000.0$+$ & \colhead{} &
\colhead{(\%)} & \colhead{(\%)} &
\colhead{(\arcdeg)} & \colhead{(\arcdeg)}}
\startdata
1996 Dec 12   & 50429.896 & 0.442 & 0.654 & 0.031 & 123.4 & 1.4\\
1997 Feb 14   & 50493.628 & 0.706 & 0.677 & 0.028 & 118.3 & 1.2\\
1997 Sep 9   & 50700.795 & 0.562 & 0.585 & 0.020 & 129.2 & 1.0\\
1997 Oct 26   & 50747.828 & 0.757 & 0.566 & 0.033 & 127.0 & 1.7\\
1997 Nov 4    & 50756.930 & 0.794 & 0.666 & 0.030 & 128.7 & 1.3\\
1998 Jan 23   & 50836.552 & 0.123 & 0.533 & 0.021 & 130.8 & 1.1\\
1998 Feb 16   & 50860.623 & 0.223 & 0.626 & 0.021 & 123.8 & 0.9\\
1998 Aug 28   & 51053.800 & 0.021 & 0.639 & 0.023 & 124.7 & 1.0\\
1999 Feb 9    & 51218.711 & 0.703 & 0.569 & 0.026 & 114.4 & 1.3\\
1999 Feb 15   & 51224.668 & 0.810 & 0.616 & 0.030 & 114.6 & 1.4\\
1999 Mar 9    & 51246.607 & 0.818 & 0.460 & 0.034 & 113.6 & 2.1
\tablenotetext{1}{Calculated with the ephemeris $2445001+ 241.9E$
(Mathieu et al. 1991).}
\enddata
\end{deluxetable}

\pagebreak \clearpage

\begin{deluxetable}{lcccccc}
\tablewidth{0pt}
\tablecaption{Polarization data for Par 1540 \label{Tab-p1540}}
\tablehead{
\colhead{UT Date} & \colhead{JD} & \colhead{Phase\tablenotemark{1}} & 
\colhead{P} & \colhead{$\sigma(P)$} &
\colhead{$\theta$} & \colhead{$\sigma(\theta)$}\\
\colhead{} & 2400000.0$+$ & \colhead{} &
\colhead{(\%)} & \colhead{(\%)} &
\colhead{(\arcdeg)} & \colhead{(\arcdeg)}}
\startdata
1996 Jan 7 & 50089.667  & 0.696  & 1.072  & 0.055    & 82.1    & 1.5\\
1996 Oct 12 & 50368.755  & 0.970  & 0.907  & 0.071    & 82.5    & 2.3\\
1997 Feb 9 & 50488.541  & 0.521  & 0.753  & 0.026    & 77.1    & 1.0\\
1997 Feb 14 & 50493.533  & 0.669  & 0.855  & 0.032    & 75.7    & 1.1\\
1997 Sep 8 & 50699.867  & 0.787  & 0.807  & 0.048    & 73.2    & 1.7\\
1997 Oct 18 & 50739.848  & 0.972  & 0.810  & 0.031    & 76.0    & 1.1\\
1997 Oct 25 & 50746.770  & 0.177  & 0.793  & 0.038    & 78.1    & 1.4\\
1997 Oct 26 & 50747.858  & 0.209  & 0.855  & 0.046    & 74.6    & 1.5\\
1997 Nov 4 & 50756.895  & 0.477  & 0.826  & 0.036    & 76.8    & 1.3\\
1998 Jan 13 & 50826.593  & 0.544  & 0.765  & 0.024    & 77.9    & 0.9\\
1998 Jan 23 & 50836.570  & 0.839  & 0.845  & 0.030    & 78.1    & 1.0\\
1998 Feb 15 & 50859.587  & 0.522  & 0.813  & 0.031    & 77.1    & 1.1\\
1998 Feb 16 & 50860.556  & 0.551  & 0.882  & 0.029    & 77.7    & 0.9\\
1998 Dec 10 & 51157.704  & 0.360  & 0.918  & 0.038    & 74.5    & 1.2\\
1999 Feb 4 & 51213.627  & 0.018  & 0.748  & 0.046    & 75.9    & 1.8\\
1999 Feb 6 & 51215.555  & 0.075  & 0.874  & 0.045    & 78.2    & 1.5\\
1999 Feb 9 & 51218.618  & 0.166  & 0.888  & 0.028    & 76.4    & 0.9\\
1999 Feb 14 & 51223.561  & 0.313  & 0.863  & 0.036    & 76.2    & 1.2\\
1999 Feb 15 & 51224.598  & 0.343  & 0.842  & 0.032    & 74.7    & 1.1\\
1999 Mar 9 & 51246.542  & 0.994  & 0.795  & 0.048    & 77.6    & 1.7
\tablenotetext{1}{Calculated with the ephemeris $2444972.95+33.73E$
(Marschall \& Mathieu 1988).}
\enddata
\end{deluxetable}

\pagebreak \clearpage

\begin{deluxetable}{lcccccc}
\tablewidth{0pt}
\tablecaption{Polarization data for Par 2486 \label{Tab-p2486}}
\tablehead{
\colhead{UT Date} & \colhead{JD} & \colhead{Phase\tablenotemark{1}} & 
\colhead{P} & \colhead{$\sigma(P)$} &
\colhead{$\theta$} & \colhead{$\sigma(\theta)$}\\
\colhead{} & 2400000.0$+$ & \colhead{} &
\colhead{(\%)} & \colhead{(\%)} &
\colhead{(\arcdeg)} & \colhead{(\arcdeg)}}
\startdata
1995 Dec 14 & 50065.738  &0.945  &0.125  &0.023  &  63.8  &  5.2\\
1996 Oct 12 & 50368.779  &0.354  &0.097  &0.039  &  59.7  & 11.6\\
1997 Feb 9 & 50488.625  &0.454  &0.117  &0.038  &  85.8  &  9.2\\
1997 Oct 19 & 50740.863  &0.072  &0.055  &0.039  &  57.4  & 20.7\\
1998 Jan 23 & 50836.613  &0.527  &0.190  &0.036  &  63.2  &  5.5\\
1998 Feb 16 & 50860.641  &0.158  &0.232  &0.036  &  58.1  &  4.5
\tablenotetext{1}{Calculated with the ephemeris $2449900.0+5.1882E$
(period from Mathieu 1994).}
\enddata
\end{deluxetable}

\pagebreak \clearpage

\begin{deluxetable}{lcccccc}
\tablewidth{0pt}
\tablecaption{Polarization data for Ori 429\label{Tab-ori429} }
\tablehead{
\colhead{UT Date} & \colhead{JD} & \colhead{Phase\tablenotemark{1}} & 
\colhead{P} & \colhead{$\sigma(P)$} &
\colhead{$\theta$} & \colhead{$\sigma(\theta)$}\\
\colhead{} & 2400000.0$+$ & \colhead{} &
\colhead{(\%)} & \colhead{(\%)} &
\colhead{(\arcdeg)} & \colhead{(\arcdeg)}}
\startdata
1995 Dec 14 & 50065.689 & 0.210 &  0.268 & 0.043 &  69.0 &  4.6\\
1996 Jan 7 & 50089.617 & 0.417 &  0.293 & 0.066 &  78.3 &  6.5\\
1996 Jan 7 & 50089.643 & 0.421 &  0.085 & 0.061 &  70.4 & 20.6\\
1996 Oct 12 & 50368.723 & 0.831 &  0.290 & 0.053 &  72.7 &  5.2\\
1997 Feb 9 & 50488.595 & 0.900 &  0.242 & 0.057 &  69.9 &  6.7
\tablenotetext{1}{Calculated with the ephemeris $2449900.0+7.460E$
(period from Mathieu 1994).}
\enddata
\end{deluxetable}

\pagebreak \clearpage

\begin{deluxetable}{lcccccc}
\tablewidth{0pt}
\tablecaption{Polarization data for Par 2494 \label{Tab-p2494}}
\tablehead{
\colhead{UT Date} & \colhead{JD} & \colhead{Phase\tablenotemark{1}} & 
\colhead{P} & \colhead{$\sigma(P)$} &
\colhead{$\theta$} & \colhead{$\sigma(\theta)$}\\
\colhead{} & 2400000.0$+$ & \colhead{} &
\colhead{(\%)} & \colhead{(\%)} &
\colhead{(\arcdeg)} & \colhead{(\arcdeg)}}
\startdata
1995 Aug 28 & 49957.860  & 0.969  & 0.352  & 0.030    & 27.8    & 2.4\\
1995 Sep 3 & 49963.800  & 0.274  & 0.178  & 0.029    & 38.9    & 4.7\\
1996 Jan 7 & 50089.715  & 0.738  & 0.178  & 0.030    & 43.1    & 4.8\\
1996 Sep 4 & 50330.868  & 0.116  & 0.231  & 0.035    & 42.9    & 4.4\\
1996 Sep 7 & 50333.844  & 0.269  & 0.124  & 0.038    & 43.0    & 8.6\\
1997 Jan 1 & 50449.724  & 0.217  & 0.117  & 0.054    & 50.3    & 13.3\\
1997 Feb 9 & 50488.569  & 0.211  & 0.149  & 0.025    & 44.1    & 4.8\\
1997 Feb 10 & 50489.566  & 0.262  & 0.114  & 0.022    & 49.6    & 5.6\\
1997 Feb 14 & 50493.508  & 0.465  & 0.121  & 0.026    & 49.2    & 6.1\\
1997 Sep 9 & 50700.859  & 0.108  & 0.190  & 0.024    & 46.0    & 3.6\\
1997 Oct 18 & 50739.870  & 0.111  & 0.186  & 0.037    & 47.8    & 5.8\\
1997 Oct 19 & 50740.844  & 0.161  & 0.203  & 0.032    & 44.8    & 4.5\\
1997 Oct 25 & 50746.797  & 0.466  & 0.159  & 0.042    & 44.9    & 7.6\\
1997 Oct 26 & 50747.894  & 0.523  & 0.191  & 0.038    & 49.5    & 5.8\\
1997 Nov 4 & 50756.847  & 0.982  & 0.121  & 0.030    & 36.4    & 7.0\\
1998 Jan 13 & 50826.546  & 0.560  & 0.194  & 0.029    & 47.2    & 4.2\\
1998 Jan 23 & 50836.592  & 0.075  & 0.148  & 0.025    & 53.6    & 4.8\\
1998 Feb 15 & 50859.562  & 0.255  & 0.212  & 0.031    & 35.3    & 4.2\\
1998 Feb 16 & 50860.527  & 0.304  & 0.191  & 0.031    & 38.5    & 4.7\\
1998 Nov 14 & 51131.644  & 0.221  & 0.224  & 0.083    & 53.5    & 10.6\\
1998 Dec 8 & 51155.797  & 0.460  & 0.237  & 0.034    & 51.2    & 4.1\\
1998 Dec 10 & 51157.725  & 0.559  & 0.162  & 0.030    & 57.6    & 5.4\\
1998 Dec 29 & 51176.758  & 0.536  & 0.107  & 0.046    & 53.5    & 12.2\\
1999 Feb 4 & 51213.604  & 0.428  & 0.121  & 0.039    & 35.7    & 9.2 \\
1999 Feb 5 & 51215.484  & 0.524  & 0.168  & 0.049    & 48.2    & 8.4\\
1999 Feb 9 & 51218.558  & 0.682  & 0.143  & 0.024    & 46.3    & 4.7\\
1999 Feb 15 & 51224.548  & 0.990  & 0.108  & 0.030    & 39.3    & 8.1\\
1999 Feb 17 & 51226.518  & 0.091  & 0.147  & 0.037    & 50.7    & 7.2\\
1999 Mar 9 & 51246.588  & 0.121  & 0.164  & 0.047    & 55.9    & 8.2\\
1999 Mar 10 & 51247.557  & 0.171  & 0.205  & 0.036    & 43.4    & 5.0
\tablenotetext{1}{Calculated with the ephemeris $2449900.0+19.4815E$
(period from Mathieu 1994).}
\enddata
\end{deluxetable}

\pagebreak \clearpage

\begin{deluxetable}{lcccccc}
\tablewidth{0pt}
\tablecaption{Polarization data for W 134 \label{Tab-w134}}
\tablehead{
\colhead{UT Date} & \colhead{JD} & \colhead{Phase\tablenotemark{1}} & 
\colhead{P} & \colhead{$\sigma(P)$} &
\colhead{$\theta$} & \colhead{$\sigma(\theta)$}\\
\colhead{} & 2400000.0$+$ & \colhead{} &
\colhead{(\%)} & \colhead{(\%)} &
\colhead{(\arcdeg)} & \colhead{(\arcdeg)}}
\startdata
1995 Dec 14 & 50065.781  & 0.108  & 0.180  & 0.040  &   26.8  &   6.4\\
1996 Jan 2 & 50084.813  & 0.104  & 0.132  & 0.052  &   11.2  &  11.3\\
1996 Jan 7 & 50089.745  & 0.880  & 0.290  & 0.061  &   36.6  &   6.0\\
1997 Jan 1 & 50449.783  & 0.550  & 0.155  & 0.067  &  -16.6  &  12.4\\
1997 Feb 9 & 50488.711  & 0.678  & 0.344  & 0.052  &   33.8  &   4.3\\
1997 Oct 19 & 50740.892  & 0.371  & 0.176  & 0.050  &   37.2  &   8.2\\
1998 Jan 23 & 50836.685  & 0.449  & 0.223  & 0.047  &   34.3  &   6.0\\
1998 Feb 16 & 50860.699  & 0.229  & 0.236  & 0.038  &   33.0  &   4.7\\
1998 Apr 29 & 50932.567  & 0.541  & 0.160  & 0.112  &  112.2  &  20.1\\
1999 Feb 9 & 51218.734  & 0.584  & 0.326  & 0.070  &   32.8  &   6.2\\
1999 Feb 16 & 51225.742  & 0.687  & 0.282  & 0.086  &   34.8  &   8.7
\tablenotetext{1}{Calculated with the ephemeris $2447472.985+6.3532E$
(Padgett \& Stapelfeldt 1994).}
\enddata
\end{deluxetable}

\pagebreak \clearpage

\begin{deluxetable}{lcccccc}
\tablewidth{0pt}
\tablecaption{Polarization data for VSB 126 \label{Tab-vsb126}}
\tablehead{
\colhead{UT Date} & \colhead{JD} & \colhead{Phase\tablenotemark{1}} & 
\colhead{P} & \colhead{$\sigma(P)$} &
\colhead{$\theta$} & \colhead{$\sigma(\theta)$}\\
\colhead{} & 2400000.0$+$ & \colhead{} &
\colhead{(\%)} & \colhead{(\%)} &
\colhead{(\arcdeg)} & \colhead{(\arcdeg)}}
\startdata
1995 Dec 14 & 50065.828  & 0.831  & 0.176  & 0.077  &   10.7  &  12.5\\
1996 Jan 2 & 50084.861  & 0.303  & 0.055  & 0.121  &   17.9  &  63.0\\
1996 Jan 7 & 50089.795  & 0.685  & 0.193  & 0.083  &  162.3  &  12.3\\
1997 Feb 9 & 50488.758  & 0.555  & 0.133  & 0.079  &  167.2  &  17.0\\
1998 Jan 23 & 50836.803  & 0.485  & 0.246  & 0.072  &   22.3  &   8.4\\
1998 Feb 16 & 50860.734  & 0.337  & 0.052  & 0.097  &   57.8  &  53.3
\tablenotetext{1}{Calculated with the ephemeris $2449900.0+12.9240E$
(period from Mathieu 1994).}
\enddata
\end{deluxetable}

\pagebreak \clearpage

\begin{deluxetable}{llcccccc}
\rotate		
\tablewidth{0pt}
\tablecaption{Polarization data for other stars \label{Tab-autres}}
\tablehead{
\colhead{Star} &
\colhead{UT Date} & \colhead{JD} & \colhead{Phase} & 
\colhead{P} & \colhead{$\sigma(P)$} &
\colhead{$\theta$} & \colhead{$\sigma(\theta)$}\\
\colhead{} & 
\colhead{} & 2400000.0$+$ & \colhead{} &
\colhead{(\%)} & \colhead{(\%)} &
\colhead{(\arcdeg)} & \colhead{(\arcdeg)}}
\startdata
UZ~Tau~E/W    & 1997 Sep 9 & 50700.684 & 0.920\tablenotemark{1}  & 0.604 & 0.035 &   47.7&    1.6\\
UZ~Tau~E/W    & 1998 Jan 23 & 50836.514 & 0.032  & 1.097 & 0.044 & 1.6&    1.1\\
DQ~Tau & 1997 Sep 9 & 50700.745 & 0.753\tablenotemark{2} & 0.571 & 0.048 & 78.9 & 2.4\\
NTTS~045251$+$3016 & 1997 Jan 1 & 50449.687 & 0.217\tablenotemark{3} & 0.026 & 0.076 &   90.9&   57.7\\
NTTS~045251$+$3016 & 1997 Oct 26 & 50747.807 & 0.335 & 0.125 & 0.050 &  119.4&   11.3\\
NTTS~045251$+$3016 & 1998 Jan 23 & 50836.535 & 0.370 & 0.095 & 0.038 &   95.0&   11.4\\
Ori~569     & 1996 Sep 7 & 50333.866 & 0.086\tablenotemark{4}  & 0.370 & 0.138 &   49.6&   10.7\\
Ori~569     & 1997 Feb 9 & 50488.653 & 0.506  & 0.250 & 0.080 &  116.9&    9.1\\
Ori~569     & 1998 Jan 23 & 50836.654 & 0.389  & 0.114 & 0.062 &   44.6&   15.8\\
Ori~569     & 1998 Feb 16 & 50860.667 & 0.039  & 0.142 & 0.086 &   38.0&   17.4
\tablenotetext{1}{Calculated with the ephemeris $2449900.0+19.1E$
(period from Mathieu et al. 1996).}
\tablenotetext{2}{Calculated with the ephemeris $2449582.54+15.8043E$
(Mathieu et al. 1997).}
\tablenotetext{3}{Calculated with the ephemeris $2449900.0+2530E$
(Mathieu 1994).}
\tablenotetext{4}{Calculated with the ephemeris $2449900.0+4.25E$ (Mathieu 1994).}
\enddata
\end{deluxetable}

\pagebreak \clearpage

\begin{deluxetable}{lcccccccc}
\tablewidth{0pt}
\tablecaption{Noise analysis and orbital inclination from the BME formalism
for some observed binaries\label{Tab-noiseBME}} 
\tablehead{
\colhead{Star} & \colhead{$DQ$} & \colhead{$\gamma$} & 
\colhead{Noise\tablenotemark{1}} & \colhead{Noise} & 
\colhead{$i({\cal O}2)$} & \colhead{$\sigma(i({\cal O}2))$} & 
\colhead{$i({\cal O}1)$} & \colhead{$\sigma(i({\cal O}1))$}\\
\colhead{} & \colhead{} & \colhead{} & 
\colhead{for $Q$} & \colhead{for $U$} & 
\colhead{(\arcdeg)} & \colhead{(\arcdeg)} & 
\colhead{(\arcdeg)} & \colhead{(\arcdeg)}}
\startdata
LkCa~3 	 & 0.346 & 4.2 & 0.25 & 0.35 & 93.9 & 25.4 & 72.3 &65.5\\
V826~Tau & 0.165 & 18.3 & 0.30 & 0.23 & 92.7 &5.3 & 25.3 & 57.7\\
GW~Ori	 & 0.109 & 42.2 & 0.36 & 0.28 & 79.7 &6.3 & 107.6 &15.7\\
Par~1540 & 0.242 &  8.5 & 0.22 & 0.17 & 96.4 &6.5 & 117.5 &38.5\\
Par~2494 & 0.140 & 25.5 & 0.19 & 0.28 & 97.3 &7.0 & 77.0 &22.9\\
W~134 	 & 0.203 & 12.1 & 0.18 & 0.25 & 83.9 &7.0 & 95.0 &22.5
\tablenotetext{1}{The noise is the square root of the variance of the fit
over the amplitude of the variations; the amplitude comes from the
maximum values of the data and not of the fit.}
\enddata
\end{deluxetable}

\pagebreak \clearpage

\begin{deluxetable}{lccccc}
\tablewidth{0pt}
\tablecaption{Other parameters returned by the BME formalism: $\Omega$, the
orientation of the orbital plane, and moments of the distribution of the
scatterers. \label{Tab-omegagammas}}
\tablehead{
\colhead{Star} & \colhead{$\Omega$} & \colhead{$\sigma(\Omega)$} &
\colhead{$\tau_0 G$} & \colhead{$\tau_0 H$} & 
\colhead{$\tau_0 G / \tau_0 H$}\\
\colhead{} & \colhead{(\arcdeg)} & \colhead{(\arcdeg)} & 
\colhead{$\times 10^{-4}$} & \colhead{$\times 10^{-4}$} & 
\colhead{}}
\startdata
LkCa~3           & 58.0  & 48.1  & 1.20 & 2.59 & 2.16\\
V826~Tau         & 96.4  & 12.6  & 4.35 & 7.32 & 1.68\\
GW~Ori           & 171.9 & 12.9  & 2.63 & 6.71 & 2.56\\
Par~1540         & 109.6 & 7.3   & 0.98 & 3.96 & 4.02\\
Par~2494         & 141.3 & 16.3  & 1.47 & 4.60 & 3.14\\
W~134            & 91.9  & 14.7  & 3.41 & 10.9 & 3.20
\enddata	   
\end{deluxetable}

\end{document}